\documentclass[aps,prx,twocolumn,amsmath,amssymb]{revtex4-2}
\usepackage{graphicx}
\usepackage[colorlinks=true,citecolor=blue,linkcolor=blue,urlcolor=blue]{hyperref}
\usepackage{amsmath,amssymb,bm,physics}
\usepackage[colorlinks=true,citecolor=blue,linkcolor=blue,urlcolor=blue]{hyperref}
\usepackage{comment}
\usepackage[caption=false]{subfig}
\usepackage{subcaption}
\usepackage{appendix}
\begin{document}

\title{Chiral electron-fluxon superconductivity in circuit quantum magnetostatics}

\author{Adel Ali}
\affiliation{Department of Physics and Astronomy, Texas A\&M University, College Station, TX, 77843 USA}
\author{Alexey Belyanin}
\affiliation{Department of Physics and Astronomy, Texas A\&M University, College Station, TX, 77843 USA}

\begin{abstract}

We investigate electron paring in two-dimensional electron systems mediated by the vacuum fluctuations of a quantized magnetic flux generated by the inductor of an LC resonator. The interaction induces long-range
attractive interactions between angular momentum states which lead to pairing in a broad class of materials with critical
temperatures of few Kelvin or even higher, depending on the field-covered area. The induced state is a pair-density wave topological chiral  superconductor.
The proposed platform in circuit QED environment offers a tunable promising tool for engineering
electron interactions in two-dimensional systems to create new quantum phases of matter.

\end{abstract}
\maketitle

\section{Introduction}
The interaction between quantized electromagnetic fields and electronic degrees of freedom has been widely studied in cavity quantum electrodynamics (QED) as a route to modifying and inducing quantum phases of matter. In that setting, cavity modes can strongly affect the electronic structure of materials and generate cavity-induced many-body phenomena \cite{Schlawin2022APR,GarciaVidal2021Sci,Lu2025AOP}. By contrast, circuit QED has been developed primarily for quantum-information applications. A lumped-element LC circuit, however, also functions as a cavity and offers a particularly versatile platform for engineering light--matter interactions \cite{nori2019}. Its capacitance, inductance, and, when present, Josephson energies can be tuned far more readily than in conventional cavity-QED architectures, providing a level of control with no direct analogue in standard cavity settings.

Here we consider charged particles coupled to the quantized magnetic flux of such a circuit cavity, namely an LC resonator. In this setting, particles exchange angular momentum through a quantized gauge field, opening a route to engineered many-body interactions and, in particular, cavity-mediated superconductivity. A key advantage of LC resonators is their extremely small effective mode volume, which yields large zero-point flux (ZPF) fluctuations and makes circuit-QED platforms especially attractive for engineering correlated electronic states. It was shown previously~\cite{interactions,magnetostatics} that quantized magnetic fields can induce such unusual interactions between charged particles; in the present work, we focus on their superconducting consequences.

Unlike conventional cavity-QED settings, the matter sector considered here lies beyond the dipole approximation, allowing orbital angular-momentum exchange between electrons and the gauge field. Because the vector potential is itself engineerable, the induced interaction can be tailored to favor specific many-body phases, including superconducting and topological states. Moreover, since the gauge field is generated by closed current loops, its spatial profile naturally acquires nontrivial angular structure. Through the Biot--Savart law, the Fourier transform of the vector potential is directly linked to the Fourier content of the underlying current distribution, so fabrication can be used to selectively enhance or suppress specific momentum components and thereby engineer interaction channels in momentum space.

Photon-mediated attractive interactions leading to superconductivity were proposed in Ref.~\cite{Schlawin2019}, where Amperean pairing between comoving electrons was predicted for a two-dimensional electron gas dispersively coupled to a nanoplasmonic cavity. However, Ref.~\cite{Andolina2024} showed that nanoplasmonic cavities do not generically provide sufficiently strong transverse current--current coupling. In the deep subwavelength regime, the interaction is instead dominated by quasi-electrostatic density--density terms, so the anticipated mode-volume compression does not generally drive the system into the Amperean-pairing regime.

Motivated by this limitation, we propose a circuit quantum magnetostatics (QMS) platform in which particles couple dispersively to a structured quantized vector potential. In contrast to conventional cavity QED, where the dominant electric-dipole coupling mediates the exchange of linear momentum, the present mechanism enables the exchange of orbital angular momentum. The resulting effective electron--electron interaction takes the form $\propto -L_iL_j$. For electrons near the Fermi surface, its magnitude scales as $v_F^2 \times $ area enclosed by the field profile, with an upper bound set by the spatial region over which the field is supported. To distinguish this mechanism from conventional Amperean pairing, we refer to the resulting instability as "flux pairing". This geometric scaling provides a route to enhancing the interaction strength and accordingly the superconducting critical temperature. The ring-cavity geometry permits multiple ($N$) resonators to couple to the same two-dimensional electron gas, thereby enhancing the field coverage of the material and effective interaction strength, with further enhancement possible through mutual inductive coupling that is beyond the scope of this work. 

In contrast to Ref.~\cite{Schlawin2019}, the present coupling is magnetostatic from the outset, so the system remains in the transverse-coupling regime even at strong coupling and the induced current--current interaction remains well defined. Unlike a static electric field, which can be represented by a scalar potential $\Phi_{\mathrm{static}}(\mathbf r)$, a static magnetic field enters through $\nabla\times\mathbf A$ and naturally mediates momentum exchange and hence current--current interactions~\cite{interactions}. In addition, magnetostatic coupling is largely insensitive to the dielectric spacer separating the superconducting ring(s) layer from the two-dimensional material; see Fig.~1.

\begin{figure}
    \centering
    \includegraphics[width=1\linewidth]{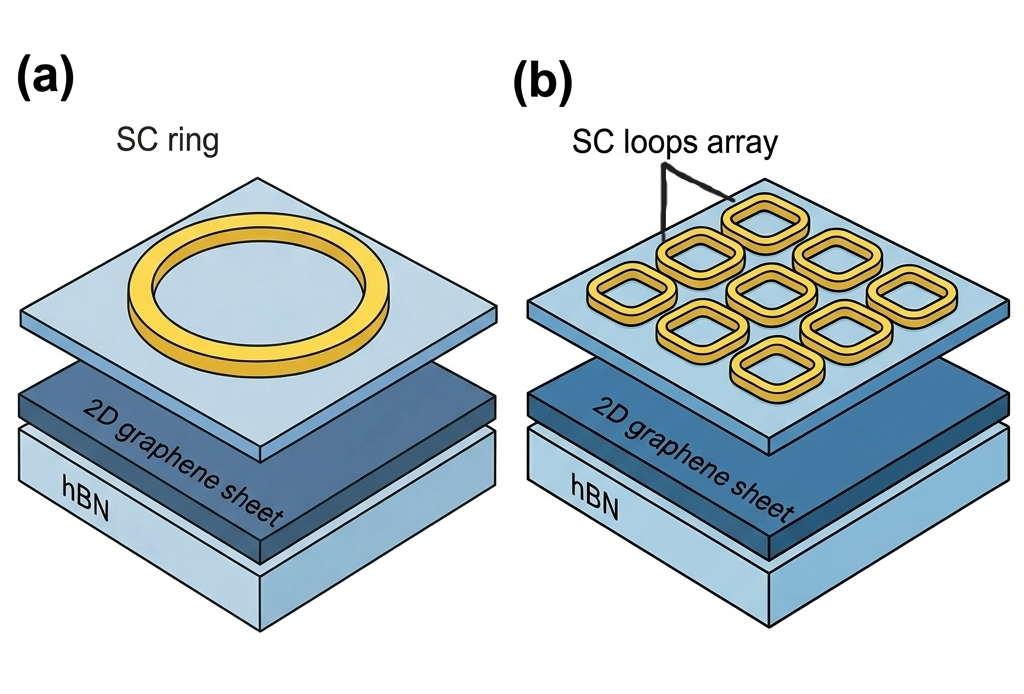}
    \caption{A sketch of an LC cavity made of a single superconducting (SC) loop (a) or an array of loops (b) which induces pairing in a 2D electron system (in this case graphene) mediated by quantum fluctuations of a quantized magnetic flux.}
    \label{fig:placeholder}
\end{figure}


 In this system, electrons interact through virtual fluxon excitations, i.e., vacuum fluctuations of the quantized gauge field. This architecture offers broad tunability of the resonator parameters $L$ and $C$. If desired, a straightforward capacitive or inductive coupling between multiple cavities, and nonlinearities controlled by the Josephson parameters $E_J$ and $E_C$ could be introduced. We show that this tunability, particularly in the inductive sector,  can drive the system into a superconducting regime.

In the dispersive regime, where the LC frequency is detuned from the relevant electronic energy scales, the cavity degree of freedom can be integrated out, yielding an emergent global interaction of the form $(\sum_i \hat O_i)^2$, where $\hat O_i$ is a electronic single-particle operator, for example, electron's angular momentum in the simplest case, $\hat O_i=\hat L_z^{(i)}$. To promote pairing, however, this operator must not be diagonal in the relevant single-particle basis. This can be achieved either by engineering an electronic band structure that breaks the symmetry associated with $\hat O$, for example through lattice geometry, external potentials, stacking, boundaries, or finite-size effects, or by introducing additional spatial structure into the vector potential itself, thereby generating nondiagonal matrix elements in the desired pairing channel.

\paragraph*{Cavity materials engineering and superconductivity.}
Cavity quantum electrodynamics (cQED) with solids has recently evolved into the broader notion of ``cavity materials engineering,'' in which the quantized electromagnetic environment acts as a control parameter for correlated phases in quantum matter \cite{Schlawin2022APR,GarciaVidal2021Sci,Hubener2024MQT,Lu2025AOP}. 
Many-body and \emph{ab initio} approaches that extend Pauli--Fierz QED to periodic solids, most notably quantum-electrodynamical density-functional theory (QEDFT) and cavity Eliashberg-type treatments, have predicted that coupling superconductors to cavity modes can renormalize electron-phonon interactions, collective modes, and even effective electron-electron interactions \cite{Sentef2018SciAdv,Schlawin2019PRL,Curtis2019PRL,Lu2024PNAS}. 
Model calculations show that cavity polaritons can enhance or suppress conventional phonon-mediated pairing depending on cavity frequency, polarization, and geometry, and may even favor unconventional pairing channels, while first-principles QEDFT simulations for MgB$_2$ find an equilibrium enhancement of $T_c$ of the order of ten percent in realistic cavity setups \cite{Sentef2018SciAdv,Lu2024PNAS}. 
Experimentally, vibrational strong coupling of molecular superconductors and cuprates to surface-plasmon or microcavity modes has been used to modify $T_c$ in both directions, consistent with an underlying cavity-controlled electron--phonon coupling \cite{Thomas2025JCP,GarciaVidal2021Sci,basov2026}. 
A recent perspective emphasizes that these developments illustrate how vacuum fluctuations inside cavities can reshape superconductivity in bulk materials, and frames cavity materials engineering as a route to equilibrium, light-controllable superconducting phases \cite{Nova2025PNAS,Lu2025AOP}. 
Taken together, these works establish cavity QED as a viable platform for engineering superconducting materials.


In a separate development on the materials side, chiral superconductors, i.e., unconventional superconductors whose order parameter has a complex phase winding in momentum space and breaks time-reversal symmetry, have emerged as a central platform for exploring topological superconductivity and Majorana excitations \cite{Kallin2016,Wysokinski2019}. The layered perovskite Sr$_2$RuO$_4$ has long been a leading candidate, with early experiments indicating unconventional pairing and spontaneous time-reversal-symmetry breaking \cite{Maeno1994,Luke1998,Xia2006}, although more recent measurements have challenged the simplest chiral $p$-wave scenario and stimulated intense debate about its actual pairing symmetry. In parallel, the heavy-fermion compound UPt$_3$ provides compelling evidence for a bulk time-reversal-symmetry-breaking state with multiple superconducting phases and possible chiral order \cite{Avers2020}, while the locally noncentrosymmetric pnictide SrPtAs shows signatures consistent with chiral $d$-wave superconductivity \cite{Biswas2013}. On the theoretical side, chiral $d+id$ states have been predicted in doped graphene near Van Hove filling and, more generally, on hexagonal lattices \cite{Nandkishore2012,BlackSchaffer2014}, and a growing body of work has identified a wide family of multicomponent and multiband superconductors in which frustration between competing pairing channels naturally generates time-reversal-symmetry-breaking condensates \cite{Kallin2016,Wysokinski2019}. 

Our proposal offers a promising platform for unconventional chiral superconductivity in ``conventional'' two-dimensional (2D) electron systems. While an applied magnetic field explicitly breaks time-reversal symmetry (TRS), the cavity vacuum has zero average field; nevertheless, cavity-mediated interactions are expected to spontaneously break TRS and yield a chiral superconducting phase. 
Depending on the microscopic regime, the structured gauge-field-mediated interaction may support either a chiral pair-density-wave state at finite momentum or a more conventional chiral superconducting state in the zero-momentum channel. Owing to the tunability of the interaction profile and lattice-scale form factors, the proposed setting may offer a controlled route for exploring the competition between these distinct superconductivity orders.

In Section II we describe the model and flux coupling. In Section III we study electron pairing on a square lattice in a uniform magnetic field over the finite area. In Section IV we generalize the treatment to a nonuniform field.   We find that in all geometries the interaction favors a chiral superconducting gap function. In Sec.~V, we present estimates of the critical temperature and discuss ways of increasing it. We also highlight the advantages of our platform relative to standard cavity QED and discuss its potential as a route toward elevated, and possibly high-temperature, superconductivity. In Section VI we show how Zeeman coupling to electron spins favors spin-triplet, chiral state in the superconducting phase. Appendix contains technical details of the derivation of some results in the main text. 



\section{Quantized vector potential of a superconducting loop}

We model a single superconducting loop as a lumped linear LC mode with canonical flux--charge pair \((\hat\Phi,\hat Q)\) obeying \([\hat\Phi,\hat Q]=i\hbar\). Quantization yields the harmonic-oscillator Hamiltonian
\begin{equation}
\hat H=\frac{\hat Q^{2}}{2C}+\frac{\hat\Phi^{2}}{2L}
=\hbar\omega\!\left(\hat a^\dagger\hat a+\tfrac{1}{2}\right), 
\qquad \omega=\frac{1}{\sqrt{LC}},
\label{eq:H-LC}
\end{equation}
with ladder-operator representation
\begin{equation}
\hat\Phi=\Phi_{\mathrm{zpf}}\left(\hat a+\hat a^\dagger\right),\quad
\hat Q=i\,Q_{\mathrm{zpf}}\left(\hat a^\dagger-\hat a\right),
\label{eq:zpf}
\end{equation}
where zero-point fluctuation amplitudes are 
$$\Phi_{\mathrm{zpf}}=\sqrt{\frac{\hbar Z}{2}},\ \ 
Q_{\mathrm{zpf}}=\sqrt{\frac{\hbar}{2Z}},$$ and \(Z=\sqrt{L/C}\) is the characteristic impedance.

For a \emph{linear} loop (no Josephson element), the loop current operator is \(\hat I=\hat\Phi/L\), so the current zero-point amplitude is
\begin{equation}
I_{\mathrm{zpf}}=\frac{\Phi_{\mathrm{zpf}}}{L}
=\sqrt{\frac{\hbar\omega}{2L}}.
\nonumber 
\end{equation}

Let \(\mathbf A_I(\mathbf r)\) be the \emph{vector potential per unit current}, defined in Coulomb gauge by the Biot--Savart line integral over the wire path \(C\),
\begin{equation}
\mathbf A_I(\mathbf r)=\frac{\mu_0}{4\pi}\oint_C \frac{d\boldsymbol\ell'}{|\mathbf r-\mathbf r'|},
\qquad 
\mathbf B_I(\mathbf r)=\nabla\times \mathbf A_I(\mathbf r),
\nonumber 
\end{equation}
so that for a static current \(I\) one has \(\mathbf A(\mathbf r)=I\,\mathbf A_I(\mathbf r)\) and \(\mathbf B(\mathbf r)=I\,\mathbf B_I(\mathbf r)\). The quantized vector-potential operator of the LC mode is therefore
\begin{equation}
\hat{\mathbf A}(\mathbf r)=I_{\mathrm{zpf}}\!\left(\hat a+\hat a^\dagger\right)\,\mathbf A_I(\mathbf r),
\;
\hat{\mathbf B}(\mathbf r)=I_{\mathrm{zpf}}\!\left(\hat a+\hat a^\dagger\right)\,\mathbf B_I(\mathbf r).
\label{eq:Aop-basic}
\end{equation}

The flux linkage through the loop satisfies
\begin{align}
\hat\Phi_{\rm loop}=\oint_C \hat{\mathbf A}\, \cdot\, d\boldsymbol\ell
=\left(\hat a+\hat a^\dagger\right)I_{\mathrm{zpf}}\oint_C \hat{\mathbf A_I}\, \cdot\,  d\boldsymbol\ell 
=L\,\hat I \nonumber \\
= \Phi_{\mathrm{zpf}}\left(\hat a+\hat a^\dagger\right),
\nonumber 
\end{align}
since \(\oint_C \mathbf A_I\!\cdot d\boldsymbol\ell=L\) by definition of the self-inductance.
The magnetic energy equals the inductive energy, 
\begin{equation}
\frac{1}{2\mu_0}\int d^3r\,|\hat{\mathbf B}|^2=\frac{1}{2}L\,\hat I^{\,2},
\nonumber 
\end{equation}
so in the vacuum, \(\langle 0| \tfrac{1}{2\mu_0}\!\int |\hat{\mathbf B}|^2 |0\rangle
=\tfrac{1}{2}L I_{\mathrm{zpf}}^2=\hbar\omega/4\), i.e., half the zero-point energy.

\paragraph*{Cavity-QED normalization.}
It is convenient to introduce a magnetoquasistatic mode function
\begin{equation}
\mathbf u_B(\mathbf r)=\frac{\mathbf A_I(\mathbf r)}{\sqrt{L}},
\qquad 
\int\!\frac{d^3r}{\mu_0}\,|\nabla\times \mathbf u_B(\mathbf r)|^2=1,
\label{eq:uBnorm}
\end{equation}
(as we will see later the effective interaction will inversely depend on $L$ so cavity compression is just reducing L)
which yields the cavity-like expressions
\begin{equation}
\hat{\mathbf A}(\mathbf r)=\sqrt{\frac{\hbar\omega}{2}}\,
\mathbf u_B(\mathbf r)\left(\hat a+\hat a^\dagger\right),
\label{eq:Aop-cavitylike}
\end{equation}
$$
\hat{\mathbf B}(\mathbf r)=\sqrt{\frac{\hbar\omega}{2}}\,
\left[\nabla\times \mathbf u_B(\mathbf r)\right]\left(\hat a+\hat a^\dagger\right).$$
These mirror the usual single-mode cavity-QED formulas in the Coulomb gauge,
\begin{equation}
\hat{\mathbf E}(\mathbf r)=i\sqrt{\frac{\hbar\omega}{2\epsilon_0}}\,
\mathbf u(\mathbf r)\left(\hat a-\hat a^\dagger\right),\,
\hat{\mathbf A}(\mathbf r)=\sqrt{\frac{\hbar}{2\epsilon_0\omega}} 
\mathbf u(\mathbf r)\left(\hat a+\hat a^\dagger\right)
\nonumber 
\end{equation}
with the electric-field normalization \(\int\epsilon_0|\mathbf u|^2=1\) replaced by the inductive normalization in Eq.~\eqref{eq:uBnorm}.

\paragraph*{Minimal coupling to matter.}
For charges \(q_i\) and masses \(m_i\) at positions \(\mathbf r_i\), the coupling is
\begin{equation}
\hat H_{\rm int}=-\sum_i\frac{q_i}{m_i}\,\hat{\mathbf p}_i\!\cdot\!\hat{\mathbf A}(\mathbf r_i)
+\sum_i\frac{q_i^{2}}{2m_i}\,\hat{\mathbf A}^2(\mathbf r_i),
\label{eq:min-coupling}
\end{equation}
with \(\hat{\mathbf A}\) from Eqs.~\eqref{eq:Aop-basic} or \eqref{eq:Aop-cavitylike}. Gauge-invariant treatments retain both the paramagnetic \(\propto\hat{\mathbf p}\!\cdot\!\hat{\mathbf A}\) and diamagnetic \(\propto\hat{\mathbf A}^2\) terms.


\section{Square lattice}
We consider spin-$\tfrac12$ fermions on a square lattice with nearest-neighbor hopping $t$ and chemical potential $\mu$.
Minimal coupling to a (quantized) vector potential is implemented by the Peierls substitution,
\begin{equation}
\hat H
=
-t\sum_{\langle ij\rangle,\sigma}
\Big(
e^{i\hat\theta_{ij}}\,\hat c_{i\sigma}^\dagger \hat c_{j\sigma}
+\text{H.c.}
\Big)
-\mu\sum_i \hat n_i
+\hbar\omega\,\hat a^\dagger \hat a,
\nonumber 
\end{equation}
where the link phase operator is
\begin{equation}
\hat\theta_{ij}
\equiv
\frac{e}{\hbar}\int_{\bm r_i}^{\bm r_j}\!\bm A(\bm r)\cdot d\bm\ell,
\quad
\bm A(\bm r)=A_\phi(r)\,\hat{\bm\phi}\,\hat X,
\nonumber 
\end{equation}
with
\begin{equation}
A_\phi(r)=A_0 f(r),
\qquad
\hat X\equiv \hat a+\hat a^\dagger.
\nonumber 
\end{equation}
Here \(\hat{\bm\phi}\) is the azimuthal unit vector and
\begin{equation}
\hat n_i=\sum_\sigma \hat c_{i\sigma}^\dagger \hat c_{i\sigma},
\nonumber 
\end{equation}
while \(f(r)\) is the radial spatial profile.

Since \(\bm A(\bm r)\propto \hat X\) commutes with fermionic bilinears, the phase factor may be written as
\begin{equation}
\hat\theta_{ij}=\alpha_{ij}\,\hat X,
\;
\alpha_{ij}\equiv \frac{eA_0}{\hbar}\int_{\bm r_i}^{\bm r_j}\! f(r)\,\hat{\bm\phi}\cdot d\bm\ell,
\;
\alpha_{ji}=-\alpha_{ij}.
\nonumber 
\end{equation}
Expanding to linear order in \(\bm A\) (i.e., keeping only the paramagnetic term),
\begin{equation}
e^{i\alpha_{ij}\hat X}\simeq 1+i\alpha_{ij}\hat X,
\nonumber 
\end{equation}
we obtain
\begin{equation}
\hat H=\hat H_{\rm el}+\hbar\omega\,\hat a^\dagger\hat a-\hat X\,\hat J,
\nonumber 
\end{equation}
with
\begin{align}
\hat H_{\rm el}
&=
-t\sum_{\langle ij\rangle,\sigma}
\big(\hat c_{i\sigma}^\dagger \hat c_{j\sigma}+\text{H.c.}\big)
-\mu\sum_i \hat n_i,
\nonumber 
\\
\hat J
&\equiv
\sum_{\langle ij\rangle}\alpha_{ij}\,\hat j_{ij},
\;
\hat j_{ij}\equiv i t\sum_\sigma
\big(\hat c_{i\sigma}^\dagger \hat c_{j\sigma}-\hat c_{j\sigma}^\dagger \hat c_{i\sigma}\big),
\nonumber 
\end{align}
where 
\(\hat j_{ij}=\hat j_{ij}^\dagger\) is the bond-current operator and \(\hat J=\hat J^\dagger\).

\paragraph*{Mode representation.}
In an arbitrary single-particle basis \(\{\varphi_\alpha(i)\}\),
\begin{equation}
\hat c_{i\sigma}=\sum_\alpha \varphi_\alpha(i)\,\hat c_{\alpha\sigma},
\nonumber 
\end{equation}
the coupling takes the bilinear form
\begin{equation}
\hat J=\sum_{\alpha,\beta,\sigma}J_{\alpha\beta}\,\hat c_{\alpha\sigma}^\dagger \hat c_{\beta\sigma},
\nonumber 
\end{equation}
with
\begin{equation}
J_{\alpha\beta}
=
it\sum_{\langle ij\rangle}\alpha_{ij}
\Big[\varphi_\alpha^\ast(i)\varphi_\beta(j)-\varphi_\alpha^\ast(j)\varphi_\beta(i)\Big].
\nonumber 
\end{equation}
Because the flux couples only to the orbital current, \(J_{\alpha\beta}\) is spin independent, within this approximation see below for spin coupling to the QMF, and satisfies
\begin{equation}
J_{\alpha\beta}=J_{\beta\alpha}^\ast.
\nonumber 
\end{equation}

\paragraph*{Dispersive elimination of the cavity.}
Consider
\begin{equation}
\hat H
=
\hat H_{\rm el}
+\hbar\omega\,\hat a^\dagger \hat a
+\hat J(\hat a+\hat a^\dagger),
\nonumber
\end{equation}
where \(\hat J=\hat J^\dagger\) acts only on the electronic sector. A purely electronic effective Hamiltonian follows directly from projection onto the displaced-oscillator ground manifold only when
\([\hat H_{\rm el},\hat J]=0\). More generally, a Schrieffer--Wolff (polaron) transformation generates additional retardation/high-frequency corrections, and the projected electronic Hamiltonian becomes
\begin{equation}
\hat H_{\rm eff}
=
\hat H_{\rm el}
-\frac{\hat J^2}{\hbar\omega}
+\frac{1}{2(\hbar\omega)^2}
[\hat J,[\hat H_{\rm el},\hat J]]
+O(\omega^{-3}).
\nonumber
\end{equation}

To isolate the two-body term, we normal order \(\hat J^2\):
\begin{align}
\hat J^{\,2}
&=
\sum_{\alpha\beta\gamma\delta}\sum_{\sigma\sigma'}
J_{\alpha\beta}J_{\gamma\delta}\,
\hat c_{\alpha\sigma}^\dagger \hat c_{\beta\sigma}
\hat c_{\gamma\sigma'}^\dagger \hat c_{\delta\sigma'}
\nonumber\\
&=
\sum_{\alpha\delta,\sigma}(J^2)_{\alpha\delta}\,
\hat c_{\alpha\sigma}^\dagger \hat c_{\delta\sigma}
-
\sum_{\alpha\beta\gamma\delta}\sum_{\sigma\sigma'}
J_{\alpha\beta}J_{\gamma\delta}\,
\hat c_{\alpha\sigma}^\dagger \hat c_{\gamma\sigma'}^\dagger
\hat c_{\beta\sigma}\hat c_{\delta\sigma'} .
\nonumber 
\end{align}
The first term on the right-hand side is a one-body renormalization and may be absorbed into \(\hat H_{\rm el}\).

In the spin triplet state, as one can see in the section below, the interaction should be written in explicitly antisymmetrized form. Expanding gives
\begin{align}
\hat H_{2{\rm b}}
& =
\frac{1}{2\hbar\omega}
\sum_{\alpha\gamma}\sum_{\beta\delta}\sum_{\sigma\sigma'}
\Big[
J_{\alpha\beta}J_{\gamma\delta}\,
\hat c_{\alpha\sigma}^\dagger
\hat c_{\gamma\sigma'}^\dagger
\hat c_{\delta\sigma'}\hat c_{\beta\sigma}
\nonumber\\
&-
J_{\alpha\delta}J_{\gamma\beta}\,
\hat c_{\alpha\sigma}^\dagger
\hat c_{\gamma\sigma'}^\dagger
\hat c_{\delta\sigma}\hat c_{\beta\sigma'}
\Big].
\nonumber 
\end{align}
This is the induced genuine two-body interaction, which is explicitly invariant under spin rotations; the spin circuit effects are discussed below. In what follows, we consider two representative radial profiles of the vector potential: a uniform magnetic profile, for which the induced interaction is highly long-ranged and nonlocal, and a Gaussian profile, for which the predominantly transverse/chiral scattering of electronic states can be demonstrated more transparently.   

\begin{figure}
    \centering
    \includegraphics[width=1\linewidth]{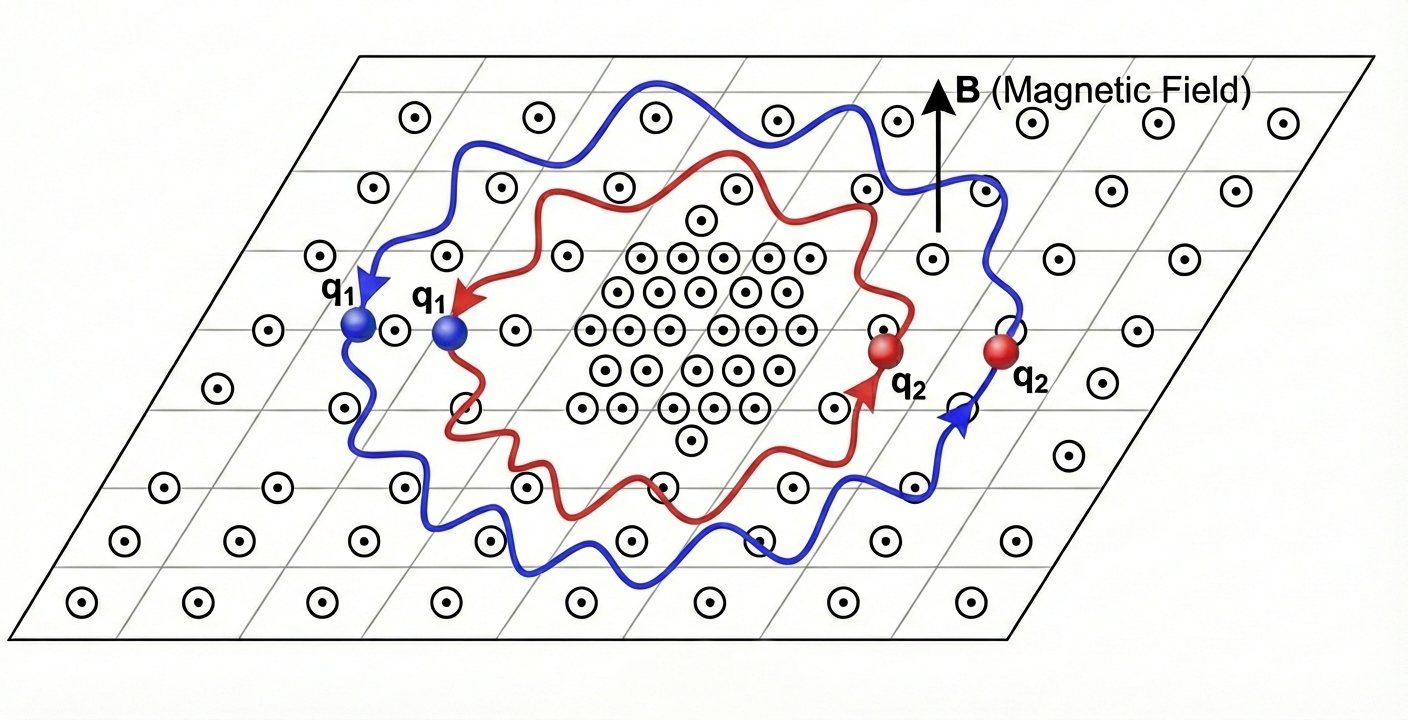}
    \caption{Two electrons encircling the quantized magnetic flux can exchange angular momentum. Unlike the conventional BCS pairing, here the pairing is between two nearby angular momentum states which breaks TRS.  }
    \label{fig:placeholder}
\end{figure}

\subsection{Uniform magnetic field}

\paragraph{Finite-$\bm Q$ spin triplet.}
We consider a single equal-spin component of a triplet state, so that the orbital
pairing problem is effectively spinless and the gap function satisfies
\begin{equation}
\Delta_{\bm Q}(-\bm k)=-\Delta_{\bm Q}(\bm k),
\label{eq:odd_PRL}
\end{equation}
where $\bm Q$ is center-of-mass pair momentum.  
After dispersive elimination of the cavity mode, the effective Hamiltonian is
\begin{equation}
\hat H_{\rm eff}
=
\sum_{\bm k}\xi_{\bm k}\,
\hat c_{\bm k}^\dagger \hat c_{\bm k}
-\frac{\hat J^2}{\hbar\omega},
\;
\xi_{\bm k}
=
-2t(\cos k_xa+\cos k_ya)-\mu ,
\nonumber 
\end{equation}
with
\begin{equation}
\hat J
=
\kappa_\Phi
\sum_i
(x_i\hat j_i^y-y_i\hat j_i^x),
\quad
\kappa_\Phi
=
\frac{ea}{2\hbar}\frac{\Phi_{\rm }}{\mathcal A}.
\nonumber 
\end{equation}
Because $\hat J$ contains $x_i,y_i$, translational invariance is broken in a finite
sample, and the pair momentum $\bm Q$ is only approximately conserved.

Writing
\begin{equation}
\hat J=\sum_{\alpha\beta}J_{\alpha\beta}\,\hat c_\alpha^\dagger \hat c_\beta ,
\nonumber 
\end{equation}
fermionic normal ordering gives
\begin{align}
-\frac{\hat J^2}{\hbar\omega}
&=
-\frac{1}{\hbar\omega}
\sum_{\alpha\delta}(J^2)_{\alpha\delta}\,
\hat c_\alpha^\dagger \hat c_\delta
\nonumber\\
&\quad+
\frac12
\sum_{\alpha\beta\gamma\delta}
V_{\alpha\beta;\gamma\delta}\,
\hat c_\alpha^\dagger \hat c_\beta^\dagger
\hat c_\delta \hat c_\gamma ,
\nonumber 
\end{align}
with the antisymmetrized two-body vertex
\begin{equation}
V_{\alpha\beta;\gamma\delta}
=
-\frac{1}{\hbar\omega}
\left(
J_{\alpha\gamma}J_{\beta\delta}
-
J_{\alpha\delta}J_{\beta\gamma}
\right).
\nonumber 
\end{equation}
The bilinear term renormalizes the one-body spectrum and is absorbed into
$\tilde\xi_{\bm k}$ below. See Appendix \ref{sec:finite_sample_matrix_elements} for the exact disscusion.

The exact finite-sample instability is obtained in the eigenbasis of the renormalized
one-body Hamiltonian,
\begin{equation}
\Delta_{nm}
=
\frac12
\sum_{rs}
V_{nm;rs}\,
\frac{1-f(\varepsilon_r)-f(\varepsilon_s)}
{\varepsilon_r+\varepsilon_s}\,
\Delta_{rs}, \,
\Delta_{mn}=-\Delta_{nm}.
\label{eq:exact_gap}
\end{equation}
To expose the leading weak-coupling structure, we project onto local plane-wave states
near a Fermi-surface patch and resolve the pairing kernel in an approximate finite
center-of-mass channel $\bm Q$:
\begin{equation}
\bm k_\pm=\frac{\bm Q}{2}\pm\bm k,
\qquad
\bm p_\pm=\frac{\bm Q}{2}\pm\bm p.
\nonumber 
\end{equation}
Here $\bm k_\pm$ and $\bm p_\pm$ are electron momenta of a pair on the Fermi surface before and after the scattering, respectively. 
The projected BCS Hamiltonian is
\begin{align}
\hat H_{\rm BCS}^{(\bm Q)}
& =
\sum_{\bm k}
\bigl(
\tilde\xi_{\bm k_+}\hat n_{\bm k_+} 
+
\tilde\xi_{\bm k_-}\hat n_{\bm k_-}
\bigr) \nonumber \\
& +
\frac{1}{2N}
\sum_{\bm k,\bm p}
V_{\bm Q}(\bm k,\bm p)\,
\hat c_{\bm k_+}^\dagger
\hat c_{\bm k_-}^\dagger
\hat c_{\bm p_-}\hat c_{\bm p_+},
\nonumber 
\end{align}
where
\begin{align}
V_{\bm Q}(\bm k,\bm p)
& \equiv
V_{\bm k_+,\bm k_-;\bm p_+,\bm p_-} \nonumber \\
& =
-\frac{1}{\hbar\omega}
\left[
J_{\bm k_+,\bm p_+}J_{\bm k_-,\bm p_-}
-
J_{\bm k_+,\bm p_-}J_{\bm k_-,\bm p_+}
\right].
\nonumber 
\end{align}
Because the vertex is already antisymmetrized,
\begin{equation}
V_{\bm Q}(-\bm k,\bm p)=-V_{\bm Q}(\bm k,\bm p),
\;
V_{\bm Q}(\bm k,-\bm p)=-V_{\bm Q}(\bm k,\bm p),
\nonumber 
\end{equation}
so the gap equation closes automatically in the odd-parity sector
\eqref{eq:odd_PRL}.

Defining
\begin{equation}
\Delta_{\bm Q}(\bm k)
=
-\frac{1}{N}
\sum_{\bm p}
V_{\bm Q}(\bm k,\bm p)\,
\langle \hat c_{\bm p_-}\hat c_{\bm p_+}\rangle ,
\label{eq:gapdef_PRL}
\end{equation}
the mean-field Hamiltonian is
\begin{align}
\hat H_{\rm MF}^{(\bm Q)}
&=
\sum_{\bm k}
\Big[
\tilde\xi_{\bm k_+}\hat c_{\bm k_+}^\dagger\hat c_{\bm k_+}
+
\tilde\xi_{\bm k_-}\hat c_{\bm k_-}^\dagger\hat c_{\bm k_-}
\nonumber\\
&\hspace{1.2cm}
+
\Delta_{\bm Q}(\bm k)\hat c_{\bm k_+}^\dagger\hat c_{\bm k_-}^\dagger
+
\text{H.c.}
\Big].
\label{eq:HMF_PRL}
\end{align}
With
\begin{equation}
\bar\xi_{\bm Q}(\bm k)
=
\frac{\tilde\xi_{\bm k_+}+\tilde\xi_{\bm k_-}}{2},
\quad
\eta_{\bm Q}(\bm k)
=
\frac{\tilde\xi_{\bm k_+}-\tilde\xi_{\bm k_-}}{2},
\nonumber 
\end{equation}
the Bogoliubov branches are
\begin{equation}
E_{\bm Q,\pm}(\bm k)
=
\eta_{\bm Q}(\bm k)
\pm
\mathcal E_{\bm Q}(\bm k),
\;
\mathcal E_{\bm Q}(\bm k)
=
\sqrt{\bar\xi_{\bm Q}^2(\bm k)+|\Delta_{\bm Q}(\bm k)|^2}.
\nonumber 
\end{equation}
The anomalous average is
\begin{equation}
\langle \hat c_{\bm k_-}\hat c_{\bm k_+}\rangle
=
-
\frac{\Delta_{\bm Q}(\bm k)}{2\mathcal E_{\bm Q}(\bm k)}
\Big[
1-f(E_{\bm Q,+})-f(E_{\bm Q,-})
\Big],
\nonumber 
\end{equation}
which yields
\begin{equation}
\Delta_{\bm Q}(\bm k)
=
\frac{1}{N}
\sum_{\bm p}
V_{\bm Q}(\bm k,\bm p)\,
\frac{\Delta_{\bm Q}(\bm p)}{2\mathcal E_{\bm Q}(\bm p)}
\Big[
1-f(E_{\bm Q,+})-f(E_{\bm Q,-})
\Big].
\label{eq:gapfull_PRL}
\end{equation}
At $T=T_c$, $\Delta_{\bm Q}\to0$, and one obtains
\begin{equation}
\Delta_{\bm Q}(\bm k)
=
\frac{1}{N}
\sum_{\bm p}
V_{\bm Q}(\bm k,\bm p)\,
\chi_{\bm Q}(\bm p;T_c)\,
\Delta_{\bm Q}(\bm p),
\label{eq:linearized}
\end{equation}
with
\begin{equation}
\chi_{\bm Q}(\bm p;T)
=
\frac{
1-f(\tilde\xi_{\bm p_+})-f(\tilde\xi_{\bm p_-})
}{
\tilde\xi_{\bm p_+}+\tilde\xi_{\bm p_-}
}.
\label{eq:chi}
\end{equation}
Equation~\eqref{eq:gapfull_PRL} is an eigenvalue problem for
$V_{\bm Q}(\bm k,\bm p)\chi_{\bm Q}(\bm p;T)$; the leading instability is the largest
eigenmode, optimized over $\bm Q$. 
To determine the preferred direction of $\bm Q$, we set
\begin{equation}
\bm Q=2\bm K,
\label{eq:Qeq2K}
\end{equation}
where $\bm K$ lies on the Fermi surface, and expand the dispersion near the corresponding patch in local normal and tangential coordinates $(q_\perp,q_\parallel)$,
\begin{align}
\xi_{\bm K+\bm q}
&=
v_F q_\perp
+\frac{\kappa_F}{2}q_\parallel^2
+\frac{q_\perp^2}{2m_\perp}
+\cdots ,
\nonumber\\
\xi_{\bm K-\bm q}
&=
-\,v_F q_\perp
+\frac{\kappa_F}{2}q_\parallel^2
+\frac{q_\perp^2}{2m_\perp}
+\cdots .
\label{eq:patch_expansion}
\end{align}
Their sum is therefore
\begin{equation}
\xi_{\bm K+\bm q}+\xi_{\bm K-\bm q}
=
\kappa_F q_\parallel^2+\frac{q_\perp^2}{m_\perp}+\cdots ,
\nonumber 
\end{equation}
and the pair susceptibility becomes sharply peaked along the tangential ridge $q_\perp=0$,
\begin{equation}
\chi_{2\bm K}(\bm q;T_c)
\sim
\frac{
1-f(\xi_{\bm K+\bm q})-f(\xi_{\bm K-\bm q})
}{
\kappa_F q_\parallel^2+q_\perp^2/m_\perp
}.
\label{eq:chi_patch}
\end{equation}
The available pairing phase space is thus largest where the local Fermi-surface curvature $|\kappa_F|$ is smallest. For the square-lattice band well inside the band, the flattest segments are generically the antinodal ones, so the dominant finite-$\bm Q$ instability is expected to occur along the axes of the lattice, 
\begin{equation}
\bm Q_\star \in
\left\{
(\pm Q_0,0),(0,\pm Q_0)
\right\},
\label{eq:Q_star}
\end{equation}
rather than along the Brillouin-zone diagonals. Equation~(\ref{eq:Q_star}) should be understood as the leading weak-coupling result within the patch approximation; the exact finite-sample kernel preserves the full $C_4$ symmetry.

The structure of the eigenfunction follows from Eq.~(\ref{eq:odd_PRL}) together with the ridge in Eq.~(\ref{eq:chi_patch}). Since the pairing weight is concentrated near $q_\perp=0$, an odd function proportional to $q_\perp$ is suppressed because it vanishes on the ridge, whereas the leading odd function supported by the ridge is odd in the tangential coordinate,
\begin{equation}
\phi_{2\bm K}(\bm q)\propto q_\parallel F(\bm q),
\qquad
F(-\bm q)=F(\bm q).
\label{eq:phi_local}
\end{equation}
Thus the leading spinless finite-$\bm Q$ state has tangential $p$-wave character around the paired patch. In lattice language, if $\bm Q_\star=(Q_0,0)$, the paired patches are centered on an antinodal region whose tangent is approximately parallel to $\hat y$, and a natural global continuation is
\begin{equation}
\phi_{(Q_0,0)}(\bm k)
\sim
g_{Q_0}(\bm k)\sin(k_y a),
\label{eq:phi_axial_x}
\end{equation}
where $g_{Q_0}(\bm k)$ is an even envelope localized near the paired patches. Likewise, for $\bm Q_\star=(0,Q_0)$ one obtains
\begin{equation}
\phi_{(0,Q_0)}(\bm k)
\sim
g_{Q_0}(\bm k)\sin(k_x a).
\label{eq:phi_axial_y}
\end{equation}
The two axial directions are degenerate at quadratic order by $C_4$ symmetry; the eventual selection between a single-$\bm Q$ stripe state and multi-$\bm Q$ combinations requires quartic terms beyond the linearized gap equation.

Finally, if the pairing kernel is projected onto its dominant odd separable channel,
\begin{equation}
V_{\bm Q}(\bm k,\bm p)
\simeq
-\lambda_{\bm Q}\,
\phi_{\bm Q}(\bm k)\phi_{\bm Q}^\ast(\bm p),
\label{eq:separable_kernel}
\end{equation}
then Eq.~(\ref{eq:linearized}) admits the analytic solution
\begin{equation}
\Delta_{\bm Q}(\bm k)=\Delta_{\bm Q}\phi_{\bm Q}(\bm k),
\label{eq:separable_gap}
\end{equation}
with scalar instability condition
\begin{equation}
1
=
\lambda_{\bm Q}\,
\frac{1}{N}\sum_{\bm p}
|\phi_{\bm Q}(\bm p)|^2
\chi_{\bm Q}(\bm p;T_c).
\label{eq:scalar_instability}
\end{equation}
The favored ordering vector is therefore maximizing the weighted susceptibility in Eq.~(\ref{eq:scalar_instability}); for the square-lattice Fermi surface deep inside the band, this selects the axial directions in Eq.~(\ref{eq:Q_star}) with tangential odd form factors given locally by Eq.~(\ref{eq:phi_local}) and globally by Eqs.~(\ref{eq:phi_axial_x}) and (\ref{eq:phi_axial_y}).

To find the ground state within this degenerate manifold, one can construct the Ginzburg-Landau (GL) free energy and evaluate the quartic (fourth-order) terms that dictate how these two modes interact. 
The resulting ground state is expected to be a chiral combination of Qs. We discuss this in more detail in the  appendix \ref{subsec:chiral_PDW_GL}.

\begin{figure*}
    \centering
    \includegraphics[width=0.9\textwidth]{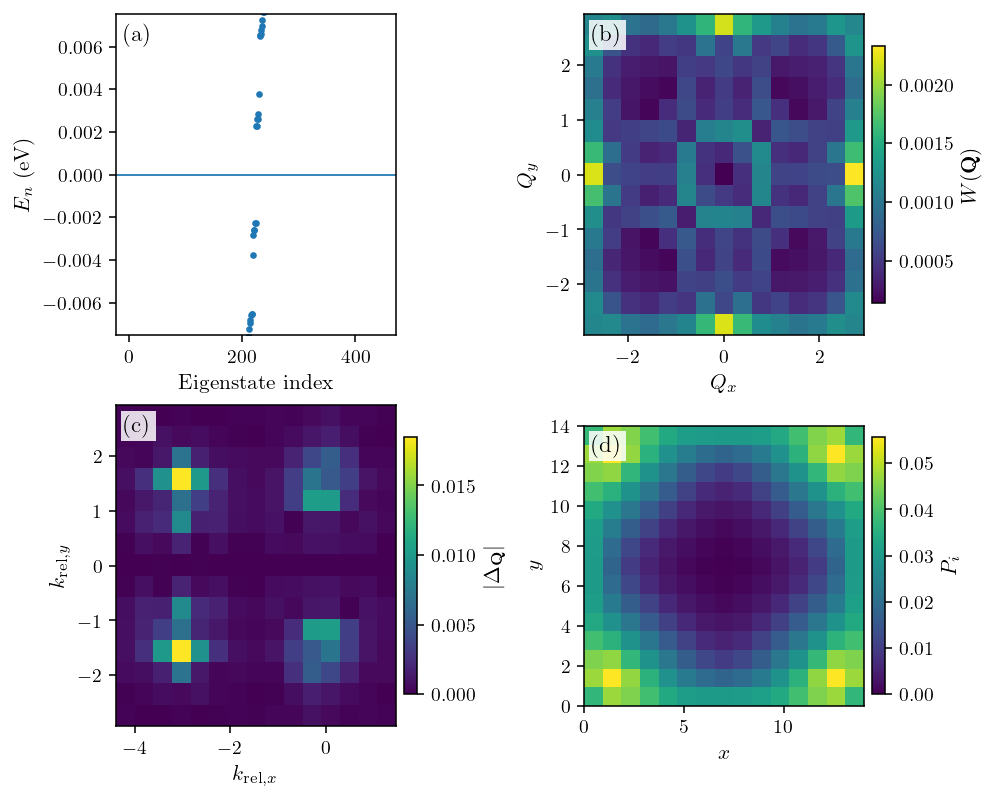}
    \caption{Mean field solution of the real space BdG Hamiltonian for the gap function. 
    (a) Low-energy BdG spectrum, showing the states nearest to zero energy. 
    (b) Center-of-mass pair-weight map $W(Q)$, demonstrating that the pairing is concentrated at finite $Q$ at axial directions. 
    (c) Magnitude of the pair amplitude $|\Delta_Q(k_{\rm rel})|$ for the dominant center-of-mass wavevector, resolved as a function of relative momentum. 
    (d) Real-space pair texture
    \(P_i=\sqrt{\sum_j |\Delta_{ij}|^2}\),
    showing the spatial distribution of the nonlocal pairing amplitude across the sample. It grows radially as expected since the interaction strength increases. 
    The figure illustrates that the solution is dominated by finite-momentum pairing with a structured internal pair wavefunction. 
   }
    \label{fig:fig2}
\end{figure*}

To characterize the cavity-mediated superconducting phase, we self-consistently solve the real-space Bogoliubov-de Gennes (BdG) equations. For spinless fermions on a two-dimensional square lattice, the mean-field Hamiltonian is expressed in the Nambu basis, $\Psi^\dagger = (c_1^\dagger, \dots, c_N^\dagger, c_1, \dots, c_N)$, as
\begin{equation}
    \mathcal{H}_{\text{MF}} = \frac{1}{2} \sum_{i,j} \begin{pmatrix} c_i^\dagger & c_i \end{pmatrix} 
    \begin{pmatrix} H_{0,ij} & \Delta_{ij} \\ -\Delta_{ij}^* & -H_{0,ij}^* \end{pmatrix} 
    \begin{pmatrix} c_j \\ c_j^\dagger \end{pmatrix},
    \label{eq:real_space}
\end{equation}
where $H_{0,ij}$, including the one-body renormalization from the field, defines the single-particle energies, and $\Delta$ is defined in Eq.~(\ref{eq:exact_gap}). As a complementary characterization of the self-consistent superconducting solution, Fig.~\ref{fig:fig2} combines spectral, center-of-mass, relative-momentum, and real-space diagnostics of the pair field. Figure~\ref{fig:fig2}(a) shows the low-energy BdG quasiparticle spectrum, highlighting the states closest to zero energy and the associated superconducting gap scale. To resolve the center-of-mass structure of the pairs, Fig.~\ref{fig:fig2}(b) displays the pair-weight distribution $W(Q)$, obtained by summing the Fourier-transformed pair amplitude over relative momentum. The pronounced maximum at finite $Q$ indicates that the superconducting order is dominated by a pair-density-wave-like component rather than uniform $Q=0$ pairing as was expected from the analytical analysis above. The internal structure of this dominant finite-momentum channel is shown in Fig.~\ref{fig:fig2}(c), which plots the magnitude of the pair amplitude $|\Delta_Q(k_{\rm rel})|$ as a function of relative momentum for the leading center-of-mass wavevector. Finally, Fig.~\ref{fig:fig2}(d) shows the real-space pair texture
\(
P_i=\sqrt{\sum_j |\Delta_{ij}|^2},
\)
revealing how the nonlocal pairing amplitude is distributed across the finite sample. Taken together, these results show that the converged superconducting state is nonlocal, finite-$Q$, and structured TRS breaking state. Increasing the coupling could make $Q=0$ the leading component. The parameters used in this calculations are similar to the ones discussed in Sec.~V below.



\section{Gaussian profile nonuniform field}

To build intuition for the nature of the interaction arising in this system, we first study a structured vector-potential profile with a Gaussian envelope. This choice provides a simple and analytically tractable model. The vector potential corresponding to a more realistic loop geometry at a finite height from the 2D plane is given in the Appendix \ref{sec:real_loop}. 

\paragraph{Square-lattice derivation from Peierls substitution.}

Consider electrons on a square lattice with lattice constant \(a\),
nearest-neighbor hopping \(t\), and a single cavity quadrature
\(\hat X\equiv \hat a+\hat a^\dagger\). This time we consider a more realistic field distribution with the cavity vector potential 
\begin{equation}
\hat{\bm A}(\bm r)=\bm A(\bm r)\,\hat X,
\qquad
\bm A(\bm r)=A_0\,(\hat{\bm z}\times \bm r)\,e^{-r^2/(2\ell^2)} ,
\label{eq:A_profile_lat}
\end{equation}
which is smooth on the scale of the lattice and decays at infinity.

The electronic Hamiltonian is introduced by Peierls substitution on each bond,
\begin{equation}
\hat H
=
-t\sum_{\langle ij\rangle,\sigma}
\left[
e^{-i\hat X\,\Phi_{ij}}
\hat c^\dagger_{i\sigma}\hat c_{j\sigma}
+\mathrm{H.c.}
\right]
-\mu\sum_{i,\sigma}\hat n_{i\sigma}
+\hbar\omega\,\hat a^\dagger\hat a ,
\label{eq:H_lat_full}
\end{equation}
where
\begin{equation}
\Phi_{ij}\equiv \frac{e}{\hbar}\int_{\bm r_i}^{\bm r_j}\bm A(\bm r)\cdot d\bm l
\nonumber 
\end{equation}
is the dimensionless Peierls phase associated with bond \(\langle ij\rangle\) and $\sigma$ is spin index.

Expanding Eq.~\eqref{eq:H_lat_full} to linear order in the cavity field gives
\begin{equation}
\hat H=\hat H_0+\hbar\omega\,\hat a^\dagger\hat a-\hat X\,\hat J+\hat H_{A^2},
\nonumber 
\end{equation}
with
\begin{equation}
\hat H_0
=
-t\sum_{\langle ij\rangle,\sigma}
\left(
\hat c^\dagger_{i\sigma}\hat c_{j\sigma}
+\mathrm{H.c.}
\right)
-\mu\sum_{i,\sigma}\hat n_{i\sigma},
\nonumber 
\end{equation}
and the paramagnetic coupling operator
\begin{equation}
\hat J
=
it\sum_{\langle ij\rangle,\sigma}
\Phi_{ij}
\left(
\hat c^\dagger_{i\sigma}\hat c_{j\sigma}
-\hat c^\dagger_{j\sigma}\hat c_{i\sigma}
\right).
\label{eq:J_lat_real}
\end{equation}
Here \(\hat H_{A^2}\) denotes the diamagnetic term, which we suppress in the following
since our focus is on the induced two-body interaction from the paramagnetic part.

\paragraph{Bonds in the nonuniform cavity.}
For nearest-neighbor bonds it is convenient to introduce directed bond variables
\(\bm\delta\in\{a\hat{\bm x},a\hat{\bm y}\}\), and bond-center coordinates
\(\bm R=\bm r_i+\bm\delta/2\). Since the field varies on the scale \(\ell\), for
\(\ell\gg a\) the Peierls phase may be evaluated in the midpoint approximation:
\begin{equation}
\Phi_{i,i+\delta}
=
\frac{e}{\hbar}\int_{\bm r_i}^{\bm r_i+\bm\delta}\bm A(\bm r)\cdot d\bm l
\simeq
\frac{ea}{\hbar}\,\bm A(\bm R)\cdot \hat{\bm\delta}.
\label{eq:midpoint_phase}
\end{equation}
For the profile in Eq.~\eqref{eq:A_profile_lat}, this gives
\begin{align}
\Phi_x(\bm R)
&\equiv \Phi_{i,i+a\hat x}
\simeq
-\Phi_0\,R_y\,e^{-R^2/(2\ell^2)}, \nonumber\\
\Phi_y(\bm R)
&\equiv \Phi_{i,i+a\hat y}
\simeq
+\Phi_0\,R_x\,e^{-R^2/(2\ell^2)},
\label{eq:Phi_xy_real}
\end{align}
where
\begin{equation}
\Phi_0\equiv \frac{eaA_0}{\hbar}.
\nonumber 
\end{equation}
Thus the lattice cavity field imprints an azimuthal bond phase with a Gaussian envelope.

\paragraph{Fourier transform of the paramagnetic operator.}
We now Fourier-transform the lattice fermions,
\begin{equation}
\hat c_{i\sigma}
=
\frac{1}{\sqrt{N}}
\sum_{\bm k}
e^{i\bm k\cdot \bm r_i}\hat c_{\bm k\sigma},
\label{eq:FT_lat}
\end{equation}
where \(N\) is the number of lattice sites. Using Eq.~\eqref{eq:J_lat_real}, and summing
over directed bonds \(\bm\delta=a\hat x,a\hat y\), we obtain
\begin{align}
\hat J
&=
it\sum_{i,\sigma}\sum_{\bm\delta}
\Phi_\delta(\bm r_i+\bm\delta/2)
\left(
\hat c^\dagger_{i\sigma}\hat c_{i+\delta,\sigma}
-\hat c^\dagger_{i+\delta,\sigma}\hat c_{i\sigma}
\right) \nonumber\\
&=
\sum_{\bm k,\bm q,\sigma}
g^{\rm lat}_{\bm k,\bm q}\,
\hat c^\dagger_{\bm k+\bm q,\sigma}\hat c_{\bm k\sigma},
\label{eq:J_kq_lat}
\end{align}
with lattice vertex
\begin{equation}
g^{\rm lat}_{\bm k,\bm q}
=-
2t\sum_{\bm\delta=a\hat x,a\hat y}
\Phi_\delta(\bm q)\,
\sin\!\left[\left(\bm k+\frac{\bm q}{2}\right)\cdot\bm\delta\right].
\label{eq:g_lat_general}
\end{equation}
Here
\begin{equation}
\Phi_\delta(\bm q)
\equiv
\frac{1}{N}\sum_{\bm R}
e^{-i\bm q\cdot \bm R}\,\Phi_\delta(\bm R)
\nonumber 
\end{equation}
is the bond-center Fourier transform of the Peierls phase.

For the Gaussian profile in Eq.~\eqref{eq:Phi_xy_real}, the Fourier transforms are odd in
\(\bm q\) and purely imaginary,
\begin{align}
\Phi_x(\bm q)
&=
+i\,\Phi_\ast\,q_y\,e^{-q^2\ell^2/2},
\nonumber\\
\Phi_y(\bm q)
&=
-i\,\Phi_\ast\,q_x\,e^{-q^2\ell^2/2},
\label{eq:Phiq_eval}
\end{align}
where \(\Phi_\ast\) is a real constant that depends on normalization conventions and is
proportional to \(\Phi_0\ell^4\). Substituting Eq.~\eqref{eq:Phiq_eval} into
Eq.~\eqref{eq:g_lat_general} yields
\begin{align}
g^{\rm lat}_{\bm k,\bm q}
=
g_\ast^{\rm lat}
\left[
q_y\sin\!\left(k_xa+\frac{q_xa}{2}\right) \right. \nonumber \\
\left. -
q_x\sin\!\left(k_ya+\frac{q_ya}{2}\right)
\right]
e^{-q^2\ell^2/2},
\label{eq:g_lat_explicit}
\end{align}
where \(g_\ast^{\rm lat}\) is purely imaginary. Equation~\eqref{eq:g_lat_explicit}
is the lattice analogue of the continuum form factor:
it is odd in momentum transfer, transverse in structure, and exponentially suppressed for
\(|\bm q|\gtrsim \ell^{-1}\).

In the long-wavelength limit \(qa\ll1\), Eq.~\eqref{eq:g_lat_explicit} reduces to
\begin{align}
g^{\rm lat}_{\bm k,\bm q}
\simeq
\tilde g_\ast\,
(\bm v_{\bm k}\times \bm q)_z\,
e^{-q^2\ell^2/2},
\nonumber \\
\bm v_{\bm k}
=
\nabla_{\bm k}\xi_{\bm k}
=
2ta(\sin k_xa,\sin k_ya),
\nonumber 
\end{align}

Thus the continuum factor \((\bm k\times\bm q)_z\) is replaced on the lattice by the
periodic band-velocity structure \((\bm v_{\bm k}\times\bm q)_z\).

\paragraph{Cavity-mediated interaction.}
Adiabatically eliminating the cavity quadrature \(\hat X\) gives the effective interaction
\begin{equation}
\hat H_{\rm int}
=
-\frac{\hat J^2}{\hbar\omega}.
\nonumber 
\end{equation}
Substituting Eq.~\eqref{eq:J_kq_lat}, and normal-ordering the fermions, we obtain the
two-body part
\begin{equation}
\hat H_{2{\rm b}}
=
-\frac{1}{\hbar\omega}
\sum_{\bm k,\bm k',\bm q}
\sum_{\sigma,\sigma'}
g^{\rm lat}_{\bm k,\bm q}\,
g^{\rm lat}_{\bm k',-\bm q}\,
\hat c^{\dagger}_{\bm k+\bm q,\sigma}
\hat c^{\dagger}_{\bm k'-\bm q,\sigma'}
\hat c_{\bm k'\sigma'}
\hat c_{\bm k\sigma},
\label{eq:H2b_lat}
\end{equation}
and suppress one-body terms that merely renormalize the dispersion.

\paragraph{Zero-momentum Cooper channel.}
if spin singlet states are allowed (see below for spin discussion), we project Eq.~\eqref{eq:H2b_lat} onto the singlet \(Q=0\) Cooper channel,
\((\bm k\uparrow,-\bm k\downarrow)\to(\bm p\uparrow,-\bm p\downarrow)\), for which
\(\bm q=\bm p-\bm k\), where $Q$ is center of mass momentum.

The corresponding BCS kernel is
\begin{equation}
V^{\rm BCS}_{\bm k,\bm p}
=
-\frac{1}{\hbar\omega}\,
g^{\rm lat}_{\bm k,\bm p-\bm k}\,
g^{\rm lat}_{-\bm k,\bm k-\bm p}.
\label{eq:VBCS_lat_def}
\end{equation}
Using Eq.~\eqref{eq:g_lat_explicit}, one finds
\begin{align}
V^{\rm BCS}_{\bm k,\bm p}
&=V_0^{\rm lat}\,
e^{-|\bm p-\bm k|^2\ell^2}
\Big[
(p_y-k_y)\sin\tfrac{(k_x+p_x)a}{2}
\nonumber\\
&\qquad\qquad
-(p_x-k_x)\sin\tfrac{(k_y+p_y)a}{2}
\Big]^2 \times
e^{-|\bm p-\bm k|^2\ell^2} .
\label{eq:VBCS_lat_exact}
\end{align}
where
\begin{equation}
V_0^{\rm lat}\equiv -\frac{(g_\ast^{\rm lat})^2}{\hbar\omega}>0,
\nonumber 
\end{equation}
since \(g_\ast^{\rm lat}\) is purely imaginary. The interaction is therefore positive
semidefinite as a kernel, but enters the gap equation with an overall minus sign.

The linearized gap equation is
\begin{equation}
\Delta_{\bm k}
=
-\frac{1}{N}\sum_{\bm p}
V^{\rm BCS}_{\bm k,\bm p}\,
\frac{\tanh(\xi_{\bm p}/2T)}{2\xi_{\bm p}}\,
\Delta_{\bm p}.
\label{eq:gap_eq_lat}
\end{equation}
Because \(V^{\rm BCS}_{\bm k,\bm p}\ge 0\), the leading pairing instability must lie in a
\emph{sign-changing} even-parity representation of the square-lattice point group.
The natural basis functions are
\begin{eqnarray}  
\eta_{A_{1g}}(\bm k) & \sim  1,\ \cos k_xa+\cos k_ya,   \nonumber \\ 
\eta_{B_{1g}}(\bm k) & =\cos k_xa-\cos k_ya, \label{eq:lattice_irreps} \\
\eta_{B_{2g}}(\bm k)& =2\sin k_xa\sin k_ya. \nonumber 
\end{eqnarray}
The \(A_{1g}\) (\(s\)-wave-like) sector is disfavored, while the sign-changing even-parity
channels \(B_{1g}\) and \(B_{2g}\) are naturally attractive.

This conclusion becomes explicit in the long-wavelength limit, where
\(\sin k_\alpha a\simeq k_\alpha a\), and Eq.~\eqref{eq:VBCS_lat_exact} reduces to
\begin{equation}
V^{\rm BCS}_{\bm k,\bm p}
\simeq
V_0\,(\bm k\times \bm p)_z^2\,
e^{-|\bm p-\bm k|^2\ell^2},
\label{eq:VBCS_cont_limit}
\end{equation}
with the standard decomposition
\begin{align}
(\bm k\times\bm p)_z^2=
(k_xp_y-k_yp_x)^2 =
\frac12(k_x^2+k_y^2)(p_x^2+p_y^2)
\nonumber\\-\frac12(k_x^2-k_y^2)(p_x^2-p_y^2)
-\frac12(2k_xk_y)(2p_xp_y).
\label{eq:cross_decomp_lat}
\end{align}
Equation~\eqref{eq:cross_decomp_lat} shows directly that the \(s\)-wave projection is
repulsive, while the \(d_{x^2-y^2}\) and \(d_{xy}\) projections are attractive.
On the square lattice, these correspond precisely to the \(B_{1g}\) and \(B_{2g}\)
representations in Eq.~\eqref{eq:lattice_irreps}.

Finally, the Gaussian factor in Eq.~\eqref{eq:VBCS_lat_exact} controls the phase space for
pairing. For \(k_F\ell\lesssim1\), large-angle scattering is substantial and the
\(d\)-wave projections remain effective. By contrast, for \(k_F\ell\gg1\) the interaction
becomes strongly forward focused, \(|\bm p-\bm k|\lesssim \ell^{-1}\ll k_F\), and the
conventional \(Q=0\) Cooper instability is weakened, opening the possibility that
finite-momentum pairing channels (or pair density wave (PDW)) may become competitive.

In this regime, the conventional \(Q=0\) Cooper instability is
substantially weakened, and finite-momentum pairing channels may become
competitive.

We now consider pairing in a finite center-of-mass channel, $\mathbf Q \neq 0,$
and focus on the spin-triplet sector.  
Because the spin
wave function is symmetric the  Fermi statistics require the orbital form factor
to be odd,
\begin{equation}
\Delta_{\mathbf Q,i}(-\mathbf k)=-\Delta_{\mathbf Q,i}(\mathbf k).
\label{eq:odd_parity_constraint}
\end{equation}

We project the interaction~\eqref{eq:H2b_lat} into the finite-$\mathbf Q$
pairing channel
\begin{equation}
\left(\frac{\mathbf Q}{2}+\mathbf k,\frac{\mathbf Q}{2}-\mathbf k\right)
\rightarrow
\left(\frac{\mathbf Q}{2}+\mathbf p,\frac{\mathbf Q}{2}-\mathbf p\right),
\end{equation}
for which the direct momentum transfer is
\begin{equation}
\mathbf q=\mathbf p-\mathbf k.
\end{equation}
The corresponding direct Cooper kernel is
\begin{equation}
V^{\rm dir}_{\mathbf Q}(\mathbf k,\mathbf p)
=
-\frac{1}{\hbar\omega}\,
g_{\mathbf Q/2+\mathbf k,\mathbf p-\mathbf k}\,
g_{\mathbf Q/2-\mathbf k,\mathbf k-\mathbf p}.
\label{eq:VQ_direct_def}
\end{equation}
Using the continuum form factor for simplicity 
\begin{equation}
g_{\mathbf{k},\mathbf q}
=
g_{\ast}\,(\mathbf{k}\times \mathbf q)_z\,e^{-q^{2}\ell^{2}/2},
\label{eq:gkq_repeat}
\end{equation}
we obtain
\begin{align}
g_{\mathbf Q/2+\mathbf k,\mathbf p-\mathbf k}
&=
g_\ast
\left[
(\mathbf k\times\mathbf p)_z
+\frac{1}{2}\bigl(\mathbf Q\times(\mathbf p-\mathbf k)\bigr)_z
\right]
e^{-|\mathbf p-\mathbf k|^2\ell^2/2}, \nonumber 
\\
g_{\mathbf Q/2-\mathbf k,\mathbf k-\mathbf p}
&=
g_\ast
\left[
(\mathbf k\times\mathbf p)_z
-\frac{1}{2}\bigl(\mathbf Q\times(\mathbf p-\mathbf k)\bigr)_z
\right]
e^{-|\mathbf p-\mathbf k|^2\ell^2/2}.
\nonumber
\end{align}
It follows that
\begin{equation}
V^{\rm dir}_{\mathbf Q}(\mathbf k,\mathbf p)
=
V_0
\left[
(\mathbf k\times\mathbf p)_z^2
-\frac{1}{4}\bigl(\mathbf Q\times(\mathbf p-\mathbf k)\bigr)_z^2
\right]
e^{-|\mathbf p-\mathbf k|^2\ell^2},
\label{eq:VQ_direct}
\end{equation}
where
\begin{equation}
V_0\equiv -\frac{g_\ast^2}{\hbar\omega}>0.
\label{eq:V0_def}
\end{equation}
Compared with the $\mathbf Q=0$ channel, Eq.~\eqref{eq:VQ_direct}
contains the new term
$-\frac14(\mathbf Q\times(\mathbf p-\mathbf k))_z^2$, which is manifestly
negative and therefore provides a direct source of attraction in the
finite-momentum channel.

For spin-triplet pairing which is stabilized by the ferromagnetic spin collective coupling (see Sec.~VI), the orbital kernel must be antisymmetrized.
The exchange contribution is
\begin{equation}
V^{\rm ex}_{\mathbf Q}(\mathbf k,\mathbf p)
=
-\frac{1}{\hbar\omega}\,
g_{\mathbf Q/2+\mathbf k,-\mathbf p-\mathbf k}\,
g_{\mathbf Q/2-\mathbf k,\mathbf p+\mathbf k},
\label{eq:VQ_exchange_def}
\end{equation}
which evaluates to
\begin{equation}
V^{\rm ex}_{\mathbf Q}(\mathbf k,\mathbf p)
=
V_0
\left[
(\mathbf k\times\mathbf p)_z^2
-\frac{1}{4}\bigl(\mathbf Q\times(\mathbf p+\mathbf k)\bigr)_z^2
\right]
e^{-|\mathbf p+\mathbf k|^2\ell^2}.
\nonumber 
\end{equation}
The full triplet pairing kernel is therefore
\begin{equation}
V^{(t)}_{\mathbf Q}(\mathbf k,\mathbf p)
=
V^{\rm dir}_{\mathbf Q}(\mathbf k,\mathbf p)
-
V^{\rm ex}_{\mathbf Q}(\mathbf k,\mathbf p),
\label{eq:triplet_kernel_def}
\end{equation}
Equation~\eqref{eq:triplet_kernel_def} makes it clear why a finite-$\mathbf Q$
instability is natural in the present problem.  The Gaussian factor favors
small momentum transfer, $|\mathbf p-\mathbf k|\lesssim \ell^{-1}$, while the
transverse structure suppresses strictly forward scattering.  At the same
time, the term proportional to $(\mathbf Q\times\mathbf q)_z^2$ generates an
explicitly attractive contribution that is absent for $\mathbf Q=0$.  This
combination is precisely the structure expected for an Amperean or
pair-density-wave-like instability.

To explore the leading odd-parity structure, it is convenient to work in a
local patch description of the Fermi surface.  Let $k_\parallel$ denote the
component parallel to $\mathbf Q$ and $k_\perp$ the component transverse to
$\mathbf Q$.  For same-patch pairing one has $k_\parallel,p_\parallel\ll k_F$,
and the dominant dependence arises from the transverse components.  In this
limit,
\begin{equation}
\bigl(\mathbf Q\times(\mathbf p\pm\mathbf k)\bigr)_z^2
\simeq
Q^2(p_\perp\pm k_\perp)^2,
\end{equation}
while the $(\mathbf k\times\mathbf p)_z^2$ term is subleading.  The triplet
kernel then reduces approximately to
\begin{align}
V^{(t)}_{\mathbf Q}(k_\perp,p_\perp)
\simeq
-\frac{V_0Q^2}{4}
\Big[
&(p_\perp-k_\perp)^2
e^{-(p_\perp-k_\perp)^2\ell^2}
\nonumber\\
&-
(p_\perp+k_\perp)^2
e^{-(p_\perp+k_\perp)^2\ell^2}
\Big].
\label{eq:patch_kernel}
\end{align}
Acting with Eq.~\eqref{eq:patch_kernel} on the odd basis function
$\phi_p(k_\perp)=k_\perp$ yields a negative eigenvalue, indicating that the
leading finite-$\mathbf Q$ triplet solution has $p$-wave character within
each patch.  In continuum notation, the corresponding form factor may be
written as
\begin{equation}
\Delta_{\mathbf Q}(\mathbf k)
\propto
(\hat{\mathbf z}\times\hat{\mathbf Q})\cdot\mathbf k.
\label{eq:pwave_formfactor}
\end{equation}
Thus, for $\mathbf Q\parallel\hat x$ one finds
$\Delta_{\mathbf Q}(\mathbf k)\propto k_y$, while for
$\mathbf Q\parallel\hat y$ one has
$\Delta_{\mathbf Q}(\mathbf k)\propto -k_x$.

The linearized gap equation in the triplet channel is
\begin{equation}
\Delta_{\mathbf Q,i}(\mathbf k)
=
-\frac{1}{N}\sum_{\mathbf p}
V^{(t)}_{\mathbf Q}(\mathbf k,\mathbf p)\,
\chi_{\mathbf Q}(\mathbf p;T)\,
\Delta_{\mathbf Q,i}(\mathbf p),
\label{eq:linearized_gap_triplet}
\end{equation}
with pair susceptibility
\begin{equation}
\chi_{\mathbf Q}(\mathbf p;T)
=
\frac{1-f(\xi_{\mathbf Q/2+\mathbf p})-f(\xi_{\mathbf Q/2-\mathbf p})}
{\xi_{\mathbf Q/2+\mathbf p}+\xi_{\mathbf Q/2-\mathbf p}}.
\label{eq:pair_susceptibility_Q}
\end{equation}
The preferred ordering wave vector $\mathbf Q$ is determined by maximizing
the leading eigenvalue of the kernel
$-V^{(t)}_{\mathbf Q}\chi_{\mathbf Q}$.

For the square-lattice dispersion, the optimal $\mathbf Q$ is selected by the geometry of the Fermi surface.
Symmetry-related candidates include the axial wave vectors
\begin{equation}
\mathbf Q=(\pm Q_0,0),\qquad (0,\pm Q_0),
\label{eq:axial_Qs}
\end{equation}
and, depending on filling, possibly diagonal vectors
$(\pm Q_0,\pm Q_0)$.  For the axial states, the corresponding odd lattice
harmonics are naturally
\begin{equation}
\Delta_{(Q_0,0)}(\mathbf k)\sim \sin(k_y a),
\qquad
\Delta_{(0,Q_0)}(\mathbf k)\sim \sin(k_x a),
\label{eq:lattice_triplet_formfactors}
\end{equation}
which are the lattice analogs of the continuum $p$-wave form
in Eq.~\eqref{eq:pwave_formfactor}.

We therefore conclude that the transverse  interaction
naturally supports a finite-momentum, spin-triplet instability with
$p$-wave structure within each $\mathbf Q$ patch, i.e., a triplet
pair-density-wave or Amperean-like superconducting state.

\subsection{Continuum limit}

\paragraph{Momentum-space vertex for the Gaussian vortex profile.}
To understand the structure of the interaction it is insightful to investigate the continuum limit where it is more analytically tractable. 
Using $\hat\psi(\bm r)=\int_{\bm k} e^{i\bm k\cdot\bm r}\hat c_{\bm k}$ with $\int_{\bm k}\equiv\int d^2k/(2\pi)^2$,
Eq.~\eqref{eq:J_lat_real} becomes
\begin{equation}
\hat J=\int_{\bm k}\int_{\bm k'} J_{\bm k\bm k'}\;\hat c_{\bm k}^\dagger \hat c_{\bm k'},
\qquad
J_{\bm k\bm k'}=\frac{e\hbar}{2m}\,(\bm k+\bm k')\cdot \tilde{\bm A}(\bm k-\bm k'),
\label{eq:Jkkp_general}
\end{equation}
where $\tilde{\bm A}(\bm q)\equiv\int d^2r\,e^{-i\bm q\cdot\bm r}\bm A(\bm r)$.
For Eq.~\eqref{eq:A_profile_lat} one finds

\begin{equation}
\tilde{\bm A}(\bm q)= -\,i\,2\pi A_0\,\ell^4\,(\hat{\bm z}\times \bm q)\,e^{-\ell^2 q^2/2},
\label{eq:Aq}
\end{equation}
and therefore
\begin{equation}
J_{\bm k\bm k'}= i\,C\,(\bm k\times \bm k')_z\,e^{-\ell^2|\bm k-\bm k'|^2/2},
\quad
C\equiv \frac{2\pi e\hbar A_0\ell^4}{m}.
\label{eq:Jkkp_final}
\end{equation}

\paragraph{Cooper channel and $p$-wave structure.}
From $\hat H_{\rm int}=-(\hbar\omega)^{-1}\hat J^2$, the antisymmetrized pairing vertex in the BCS channel ($\bm Q=0$) is
\begin{equation}
V(\bm k,\bm k')=\frac{1}{\hbar\omega}\Big[J_{\bm k\bm k'}J_{-\bm k,-\bm k'}-J_{\bm k,-\bm k'}J_{-\bm k,\bm k'}\Big].
\nonumber 
\end{equation}
Using Eq.~\eqref{eq:Jkkp_final} gives the explicit kernel
\begin{equation}
V(\bm k,\bm k')
=
-\frac{C^2}{\hbar\omega}\,(\bm k\times \bm k')_z^{\,2}\,
\Big[e^{-\ell^2|\bm k-\bm k'|^2}-e^{-\ell^2|\bm k+\bm k'|^2}\Big].
\nonumber 
\end{equation}
Projecting to the Fermi surface $|\bm k|=|\bm k'|=k_F$ implies that $V$ depends only on the relative angle
$\Delta\theta=\theta-\theta'$; hence the linearized gap equation
\begin{equation}
\Delta(\theta)=-N_0\int\frac{d\theta'}{2\pi}\,V(\theta-\theta')\,\Delta(\theta')\;\ln\!\frac{\Lambda}{T}
\label{eq:gap_linearized}
\end{equation}
is diagonal in angular momentum, with eigenfunctions $\Delta_\ell(\theta)\propto e^{i\ell\theta}$.
For spinless fermions, antisymmetry requires $\Delta(-\bm k)=-\Delta(\bm k)$, i.e.\ $\ell$ odd.
The leading instability therefore occurs in the lowest odd channel,
\begin{equation}
\Delta(\bm k)\propto e^{i\theta_{\bm k}}
\ \ \Longleftrightarrow\ \ 
\Delta(\bm k)\propto (k_x+i k_y)
\nonumber 
\end{equation}
on/near the Fermi surface, corresponding to chiral $p$-wave pairing (or its real $p_x/p_y$ combinations).

For the finite pair momentum $Q$,  
projecting to the Fulde--Ferrell (FF) pairing manifold with center-of-mass momentum $\bm Q$,
we define $\Delta(\bm k)\equiv \Delta_{\bm k+\bm Q/2,\,-\bm k+\bm Q/2}$ and obtain the linearized
gap equation near $T_{c}$,
\begin{equation}
\Delta(\bm k)= -\sum_{\bm k'} V^{(\bm Q)}_{\bm k,\bm k'}\,\chi_{\bm k'}(T)\,\Delta(\bm k'),
\nonumber 
\end{equation}
where $\chi_{\bm k}(T)$ is the pair susceptibility and the Cooper kernel reads
\begin{align}
V^{(\bm Q)}_{\bm k,\bm k'}
& =
\frac{1}{\hbar\omega}
\Big[
J_{\bm k+\bm Q/2,\ \bm k'+\bm Q/2}\,
J_{-\bm k+\bm Q/2,\ -\bm k'+\bm Q/2} \nonumber \\
& -
J_{\bm k+\bm Q/2,\ -\bm k'+\bm Q/2}\,
J_{-\bm k+\bm Q/2,\ \bm k'+\bm Q/2}
\Big].
\nonumber 
\end{align}

In the regime $a\equiv (k_{F}\ell)^{2}\gg 1$ the kernel is dominated by near-forward scattering
$\bm k'=\bm k+\bm\delta$ with $|\bm\delta|\ll k_{F}$. On the Fermi surface, writing
$\bm k=k_{F}(\cos\theta,\sin\theta)$, $\bm Q=Q(\cos\theta_{Q},\sin\theta_{Q})$, and taking
$\bm\delta$ tangential to the Fermi circle ($\bm k\cdot \bm\delta\simeq 0$), we find
\begin{align}
V^{(\bm Q)}(\theta,\theta+\delta\theta) & \simeq
- g_{0}\,k_{F}^{4}\,\delta\theta^{2}\,e^{-a\,\delta\theta^{2}}
\left[
1-\frac{Q^{2}}{4k_{F}^{2}}\cos^{2}(\theta-\theta_{Q})
\right], \nonumber \\
g_{0} & \equiv \frac{C^{2}}{\hbar\omega}.
\label{eq:VQ_forward}
\end{align}
The angular dependence in \eqref{eq:VQ_forward} can be decomposed as
\begin{equation}
1-\frac{Q^{2}}{4k_{F}^{2}}\cos^{2}(\theta-\theta_{Q})
=
\left(1-\frac{Q^{2}}{8k_{F}^{2}}\right)
-\frac{Q^{2}}{8k_{F}^{2}}\cos\!\big[2(\theta-\theta_{Q})\big].
\nonumber 
\end{equation}
The first term preserves rotational invariance and admits angular-momentum eigenfunctions
$\Delta(\theta)\propto e^{i\ell\theta}$ (odd $\ell$ for spinless fermions).
The second term explicitly breaks continuous rotational symmetry down to a twofold anisotropy
and mixes harmonics differing by $\Delta \ell=\pm 2$:
\begin{equation}
e^{i\ell\theta}\cos\!\big[2(\theta-\theta_{Q})\big]
=
\frac12\left(
e^{-2i\theta_{Q}}\,e^{i(\ell+2)\theta}
+
e^{+2i\theta_{Q}}\,e^{i(\ell-2)\theta}
\right).
\nonumber 
\end{equation}
Consequently, the leading odd-parity solution (dominantly $\ell=1$ for $a\gg 1$) acquires a
controlled admixture of $\ell=3$ at order $Q^{2}/k_{F}^{2}$,
\begin{equation}
\Delta(\theta)\approx
\Delta_{1}\,e^{i\theta}
+
\alpha(a,T)\,\frac{Q^{2}}{k_{F}^{2}}\,
\Delta_{1}\,e^{-2i\theta_{Q}}\,e^{i3\theta}
+O\!\left(\frac{Q^{4}}{k_{F}^{4}}\right),
\nonumber 
\end{equation}
where $\alpha(a,T)=O(1)$ is set by the ratio of the $\ell=3$ and $\ell=1$ eigenvalues of the
forward-scattering kernel (and thus depends parametrically on $a$ and on the thermal shell
encoded in $\chi_{\bm k}(T)$). In momentum components this corresponds to a predominantly
chiral $p$-wave form with a small $f$-wave correction aligned with $\bm Q$,
\begin{align}
\Delta(\bm k)\ & \propto\ (k_{x}+ik_{y})
+\alpha(a,T)\,\frac{Q^{2}}{k_{F}^{2}}\,e^{-2i\theta_{Q}}\,(k_{x}+ik_{y})^{3}
+\cdots, \nonumber \\
|\bm k| & = k_{F}.
\nonumber 
\end{align}
In real space the FF order parameter carries the center-of-mass modulation
$\Delta(\bm r)\sim e^{i\bm Q\cdot \bm r}$.

\section{interaction Energy and critical temperature estimation }

In our estimates, for a $B_{zpf}$ field covering an area A, the effective interaction energy scales as
\begin{equation}
E_{\mathrm{int}}
\simeq \frac{q^{2} B_{0}^{2}}{4\hbar\Omega}\,L_{z}^{2}
\simeq \frac{q^{2} B_{0}^{2} v_{F}^{2}}{4\hbar\Omega}\,A
\equiv \alpha\,A ,
\nonumber 
\end{equation}
where $A$ is the sample area and $v_{F}$ is the Fermi velocity; here we take graphene.  For a representative set of parameters,
$q=e$, $B_{0}=1~\mathrm{mT}$, $v_{F}=10^{6}~\mathrm{m/s}$, and $\Omega/2\pi=0.5~\mathrm{THz}$, the prefactor evaluates to
\begin{equation}
\alpha
= \frac{q^{2} B_{0}^{2} v_{F}^{2}}{4\hbar\Omega}
\approx 1.9\times 10^{-11}~\mathrm{J/m^{2}}
\approx 1.2\times 10^{-4}~\mathrm{eV/\mu m^{2}} .
\nonumber 
\end{equation}

$L_z$ is quantized so the estimate has to be rounded to an integer. 

The weak-coupling BCS relation for the critical temperature reads  $$2\Delta(0) \approx 3.53\, k_B T_c. $$ Investigating the solution of Eq.~(\ref{eq:real_space}), this corresponds to $T_c \sim 1$ K for 1 $\mu m^2$ area.

This mechanism may offer a promising route towards enhanced superconducting critical temperatures and even high temperature SC . Because the pairing interaction is mediated by quantized magnetic field, extending its spatial coverage over a larger portion of the two-dimensional material can strengthen the superconducting instability, as one can see that the interaction energy scales with the magnetic flux. In contrast to conventional cQED platforms, which are typically limited to a single mode whose field amplitude diminishes with increasing mode volume, the present geometry can be scaled by using multiple loops to cover a larger area, as shown in Fig.~1(b). This form of geometric extensibility is difficult to realize in standard cQED setups.

Expanding the area over which the magnetic field is supported increases the interaction energy until the coupling becomes sufficiently strong to excite the circuit and drive the system out of the dispersive regime. In this strong-coupling limit, the approximation of an instantaneous interaction ceases to be valid, and the effective coupling is mediated by real circuit excitations.  

in the presented model,the circuit is a dark state, vacuum $| n\rangle=0$, and the fluctuations of the vacuum flux state mediate the interactions. Higher $| n\rangle$ Fock states that are easier to implement in the circuit QED setting than in the cQED will enhance the interaction \cite{Chakraborty}. 
Compared with atomic cQED, circuit QED offers a broader and more engineerable route to the preparation of large Fock states. The key advantage is the possibility of adding Josephson nonlinearity together with in situ dc and flux-bias control, which substantially enlarges the accessible control space for state stabilization \cite{Deng2024LargeFockStates}.
Since $\langle \hat B \rangle=0$ in a Fock state, the quantized field does not break the TRS 
unlike coherent states that will have nonzero expectation value so that the interaction and the system will be closer to the classical magnetic field.

Experimentally, this phase may be identified through its magneto-optical response, which provides a direct probe of time-reversal-symmetry breaking. Moreover, if the superconducting state is indeed triplet, the emergence of half-quantum vortices is expected.


\section{Zeeman coupling to spin}

Graphene is known for having small spin-orbit coupling. However, as we show in this section,  the cavity can mediate total spin-orbit coupling or locking. Also the cavity can induce ferromagnetic interactions for the spins. 

We now include spin degrees of freedom and let the quantized magnetic mode
couple both to the orbital motion and to the spin. The electron spin couples to the same quantized magnetic field
via a Zeeman term
\begin{equation}
\hat H_{\mathrm{spin}}
= -\mu_B B_0\big(\hat a + \hat a^\dagger\big)
\sum_{i=1}^N \hat\sigma_i^z,
\label{eq:H_spin_mode}
\end{equation}
where $\mu_B$ is the Bohr magneton (we absorb the $g$-factor into $\mu_B$).

It is convenient to separate purely electronic and bosonic parts. We introduce the orbital and spin operators
\begin{align}
\hat F_{\mathrm{orb}} &= -\frac{q B_0}{m}\sum_{i=1}^N \hat L_z^{(i)}, \nonumber \\
\hat F_{\mathrm{spin}} &= -\mu_B B_0\sum_{i=1}^N \hat\sigma^{(i)}_z. \nonumber 
\end{align}
The electron--mode coupling then takes the compact form
\begin{equation}
\hat H_{\mathrm{int}} = (\hat a + \hat a^\dagger)\,\hat X,
\quad
\hat X \equiv \hat F_{\mathrm{orb}} + \hat F_{\mathrm{spin}},
\nonumber 
\end{equation}
so that the full Hamiltonian is
\begin{equation}
\hat H
= \hat H_e + \hat H_{\mathrm{mode}} + (\hat a + \hat a^\dagger)\hat X.
\nonumber 
\end{equation}

In the dispersive regime, where the field mode is far detuned from electronic
excitations and remains close to its ground state,
\begin{equation}
\|\hat X\|\ll \hbar\Omega,
\nonumber 
\end{equation}
we can integrate out the bosonic degree of freedom perturbatively (e.g.,
via a Schrieffer--Wolff transformation). To second order in $\hat X$ one
obtains an effective purely electronic Hamiltonian
\begin{equation}
\hat H_{\mathrm{eff}}
= \hat H_e - \frac{1}{\hbar\Omega}\,\hat X^2
+ \mathcal{O}\!\left(\frac{\hat X^3}{(\hbar\Omega)^2}\right),
\nonumber 
\end{equation}
up to an additive constant. Using $\hat X = \hat F_{\mathrm{orb}}+\hat F_{\mathrm{spin}}$,
we have
\begin{equation}
\hat X^2
= \hat F_{\mathrm{orb}}^2 + \hat F_{\mathrm{spin}}^2
+ 2\,\hat F_{\mathrm{orb}}\hat F_{\mathrm{spin}},\nonumber 
\end{equation}
so that
\begin{equation}
\hat H_{\mathrm{eff}}
= \hat H_e
- \frac{1}{\hbar\Omega}\hat F_{\mathrm{orb}}^2
- \frac{1}{\hbar\Omega}\hat F_{\mathrm{spin}}^2
- \frac{2}{\hbar\Omega}\hat F_{\mathrm{orb}}\hat F_{\mathrm{spin}}.
\nonumber 
\end{equation}

The first interaction term is orbit-orbit coupling seen before,
\begin{equation}
-\frac{1}{\hbar\Omega}\hat F_{\mathrm{orb}}^2
= -\frac{q^2 B_0^2}{m^2\hbar\Omega}
\left(\sum_{i=1}^N \hat L_z^{(i)} \right)^2,
\nonumber 
\end{equation}
which is a current--current interaction in the orbital sector. In momentum space and
projected to the Fermi surface it generates a separable attractive kernel whose leading eigenfunction
is odd under $k_y\to -k_y$. i.e., a superconducting instability in the $p$-wave orbital channel.

The second term,
\begin{equation}
-\frac{1}{\hbar\Omega}\hat F_{\mathrm{spin}}^2
= -\frac{\mu_B^2 B_0^2}{\hbar\Omega}
\left(\sum_{i=1}^N \hat\sigma_i^z\right)^2,
\nonumber 
\end{equation}
describes an infinite-range Ising interaction between spins. Expanding
$\left(\sum_i \hat\sigma_i^z\right)^2 = N + 2\sum_{i<j}\hat\sigma_i^z\hat\sigma_j^z$,
we see that the nontrivial part is a ferromagnetic coupling
\begin{equation}
\hat H_{\mathrm{spin\text{-}eff}}
= -J_s \sum_{i<j}\hat\sigma_i^z\hat\sigma_j^z,
\quad
J_s = \frac{2\mu_B^2 B_0^2}{\hbar\Omega} > 0,
\nonumber 
\end{equation}
which tends to polarize the electronic spins along the $z$-axis. In the
spin-polarized regime the superconducting instability is necessarily
\emph{triplet}, involving equal-spin pairs (e.g., $\uparrow\uparrow$).

Finally, the cross term
\begin{equation}
-\frac{2}{\hbar\Omega}\hat F_{\mathrm{orb}}\hat F_{\mathrm{spin}}
= -\frac{2q\mu_B B_0^2}{m\hbar\Omega}
\left(\sum_{i=1}^N \hat L_z^{(i)}\right)
\left(\sum_{j=1}^N \hat\sigma_j^z\right)
\nonumber 
\end{equation}
provides a magneto-orbital coupling that locks the direction of the orbital
current to the spin polarization. In a mean-field description, a finite
$\langle \sum_j \hat\sigma_j^z\rangle$ acts as an effective field favoring a
particular sign of $\langle\sum_i \hat x^{(i)}\hat p_y^{(i)}\rangle$, thereby
selecting the chirality or orientation of the chiral state and stabilizing the spin triplet.

In summary, integrating out the magnetic mode generates (i) an orbital
current--current interaction that drives a $p$-wave superconducting instability,
(ii) a ferromagnetic Ising interaction that spin-polarizes the electrons,
and (iii) a cross term that couples spin polarization to orbital chirality.
The resulting superconducting phase is naturally a \emph{spin-triplet,
chiral} state, built from equal-spin Cooper pairs on a polarized Fermi
surface.

\section{conclusion}

In conclusion, our results identify the proposed platform as a promising and potentially realizable route to different unconventional superconductivity phases of matter. Depending on the microscopic regime, the same structured gauge-field-mediated interaction can favor either chiral pair-density-wave order at finite momentum or a uniform chiral superconducting state in the zero-momentum channel. More broadly, the high degree of tunability in the interaction profile and pairing structure makes this setting a compelling arena for engineering and probing competing chiral superconducting phases in a controlled manner.

Chiral superconductivity was recently observed in rhombohedral graphene. Our results could provide an explanation if at least one of the layers contain magnetic impurities that could facilitate a similar type of angular momentum pairing.  

Our results highlight a qualitative distinction from standard cavity QED: the relevant coupling here is not electric dipole, but to the orbital motion, thereby enabling angular-momentum exchange between electrons. The induced interaction, proportional to $-L_iL_j$, grows for Fermi-surface electrons as $v_F^2$ times the field-covered area. Increasing the spatial extent of the supported magnetic field can therefore enhance the interaction energy scale and, in turn, provide a favorable route toward higher critical temperatures.

\appendix

\section{Finite-sample interaction matrix elements for uniform magnetic field}
\label{sec:finite_sample_matrix_elements}

We consider spinless fermions on a finite $L\times L$ square lattice with open boundary conditions,
\begin{equation}
\hat H_e
=
-t\sum_{\langle ij\rangle}
\left(
\hat c_i^\dagger \hat c_j+\mathrm{H.c.}
\right)
-\mu\sum_i \hat n_i ,
\label{eq:He_finite}
\end{equation}
and the cavity-induced effective Hamiltonian
\begin{equation}
\hat H_{\rm eff}
=
\hat H_e
-\frac{\hat J^2}{\hbar\omega}.
\label{eq:Heff_finite}
\end{equation}
For a uniform magnetic field in symmetric gauge,
\begin{equation}
\mathbf A(\mathbf r)
=
\frac{B}{2}(-y,x,0)\,\hat X,
\qquad
\hat X\equiv \hat a+\hat a^\dagger ,
\label{eq:A_symmetric_finite}
\end{equation}
the electronic operator coupled to the cavity is
\begin{equation}
\hat J
=
\kappa_\Phi
\left[
\sum_{x=1}^{L}\sum_{y=1}^{L-1}
X_x\,\hat j_y(x,y)
-
\sum_{x=1}^{L-1}\sum_{y=1}^{L}
Y_y\,\hat j_x(x,y)
\right],
\nonumber 
\end{equation}
where
\begin{equation}
\kappa_\Phi
=
\frac{ea}{2\hbar}\frac{\Phi}{\mathcal A},
\label{eq:kappaPhi_finite}
\end{equation}
and we measure position from the sample center,
\begin{equation}
X_x=a\left(x-\frac{L+1}{2}\right),
\qquad
Y_y=a\left(y-\frac{L+1}{2}\right).
\label{eq:centered_coordinates}
\end{equation}
The bond-current operators are
\begin{align}
\hat j_x(x,y)
&=
it\left(
\hat c_{x+1,y}^\dagger \hat c_{x,y}
-
\hat c_{x,y}^\dagger \hat c_{x+1,y}
\right),
\label{eq:jx_def}
\\
\hat j_y(x,y)
&=
it\left(
\hat c_{x,y+1}^\dagger \hat c_{x,y}
-
\hat c_{x,y}^\dagger \hat c_{x,y+1}
\right).
\label{eq:jy_def}
\end{align}

Let $\{\phi_\alpha(i)\}$ denote the normalized single-particle eigenstates of
$\hat H_e$,
\begin{equation}
\sum_j h_{ij}\phi_\alpha(j)=\varepsilon_\alpha \phi_\alpha(i),
\qquad
\hat c_i=\sum_\alpha \phi_\alpha(i)\hat c_\alpha ,
\label{eq:eigenbasis_def}
\end{equation}
so that
\begin{equation}
\hat H_e=\sum_\alpha \varepsilon_\alpha \hat c_\alpha^\dagger \hat c_\alpha .
\label{eq:He_diagonal}
\end{equation}
In this basis,
\begin{equation}
\hat J=\sum_{\alpha\beta}J_{\alpha\beta}\,
\hat c_\alpha^\dagger \hat c_\beta ,
\label{eq:J_matrix_basis}
\end{equation}
with matrix elements
\begin{align}
J_{\alpha\beta}
&=
it\,\kappa_\Phi
\Bigg[
\sum_{x=1}^{L}\sum_{y=1}^{L-1}
X_x\,\Lambda^{(y)}_{\alpha\beta}(x,y)
-
\sum_{x=1}^{L-1}\sum_{y=1}^{L}
Y_y\,\Lambda^{(x)}_{\alpha\beta}(x,y)
\Bigg],
\label{eq:Jalphabeta_general}
\end{align}
where 
\begin{align}
\Lambda^{(y)}_{\alpha\beta}(x,y)
&\equiv
\phi_\alpha^*(x,y+1)\phi_\beta(x,y)
-
\phi_\alpha^*(x,y)\phi_\beta(x,y+1),
\label{eq:Lambda_y}
\nonumber
\\
\Lambda^{(x)}_{\alpha\beta}(x,y)
&\equiv
\phi_\alpha^*(x+1,y)\phi_\beta(x,y)
-
\phi_\alpha^*(x,y)\phi_\beta(x+1,y).
\nonumber 
\end{align}

This is the exact finite-sample matrix element of the angular-momentum . Because $X_x$ and $Y_y$ appear explicitly, translational invariance is lost, and crystal momentum is no longer a good quantum number.

Fermionic normal ordering of $-\hat J^2/(\hbar\omega)$ yields
\begin{align}
-\frac{\hat J^2}{\hbar\omega}
&=
-\frac{1}{\hbar\omega}
\sum_{\alpha\delta}
(J^2)_{\alpha\delta}\,
\hat c_\alpha^\dagger \hat c_\delta
+
\frac12
\sum_{\alpha\beta\gamma\delta}
V_{\alpha\beta;\gamma\delta}\,
\hat c_\alpha^\dagger \hat c_\beta^\dagger
\hat c_\delta \hat c_\gamma ,
\nonumber 
\end{align}
where
\begin{equation}
V_{\alpha\beta;\gamma\delta}
=
-\frac{1}{\hbar\omega}
\left(
J_{\alpha\gamma}J_{\beta\delta}
-
J_{\alpha\delta}J_{\beta\gamma}
\right)
\label{eq:V_abgd}
\end{equation}
is the antisymmetrized two-body vertex. The effective one-body sector is therefore
\begin{equation}
\hat H_{\rm 1b}^{\rm eff}
=
\sum_{\alpha\beta}
h^{\rm eff}_{\alpha\beta}\,
\hat c_\alpha^\dagger \hat c_\beta,
\quad
h^{\rm eff}_{\alpha\beta}
=
\varepsilon_\alpha\delta_{\alpha\beta}
-\frac{(J^2)_{\alpha\beta}}{\hbar\omega}.
\nonumber 
\end{equation}

In a finite sample, $(J^2)_{\alpha\beta}$ is not generally diagonal in the eigenbasis of $\hat H_e$. Thus, the cavity-induced bilinear term cannot in general be absorbed into a simple scalar shift of $\varepsilon_\alpha$; rather, one must first diagonalize the full matrix $h^{\rm eff}_{\alpha\beta}$.

For the $L\times L$ square with open boundaries, the eigenstates of $\hat H_e$ factorize as
\begin{equation}
\phi_{\mathbf n}(x,y)
=
u_{n_x}(x)u_{n_y}(y),
u_n(x)=\sqrt{\frac{2}{L+1}}\sin\!\left(\frac{\pi n x}{L+1}\right),
\nonumber 
\end{equation}
with $\mathbf n=(n_x,n_y)$ and eigenvalues
\begin{equation}
\varepsilon_{\mathbf n}
=
-2t
\left[
\cos\!\left(\frac{\pi n_x}{L+1}\right)
+
\cos\!\left(\frac{\pi n_y}{L+1}\right)
\right]
-\mu .
\nonumber 
\end{equation}
In this basis, Eq.~\eqref{eq:Jalphabeta_general} reduces to
\begin{equation}
J_{\mathbf n\mathbf m}
=
it\,\kappa_\Phi
\left[
X_{n_xm_x}\,D_{n_ym_y}
-
D_{n_xm_x}\,X_{n_ym_y}
\right],
\label{eq:Jnm_factorized}
\end{equation}
where
\begin{align}
X_{nm}
&=
\sum_{x=1}^{L}
X_x\,u_n(x)u_m(x),
\nonumber 
\\
D_{nm}
&=
\sum_{x=1}^{L-1}
\left[
u_n(x+1)u_m(x)-u_n(x)u_m(x+1)
\right].
\nonumber 
\end{align}
These matrix elements obey the parity selection rule $X_{nm}=D_{nm}=0$ for $n+m$ even, while for $n+m$ odd one finds
\begin{align}
D_{nm}
&=
\frac{4}{L+1}
\frac{\sin k_n\,\sin k_m}{\cos k_n-\cos k_m},
\label{eq:Dnm_closed}
\\
X_{nm}
&=
-\frac{2a}{L+1}
\frac{\sin k_n\,\sin k_m}{(\cos k_n-\cos k_m)^2},
\label{eq:Xnm_closed}
\end{align}
with
\begin{equation}
k_n=\frac{\pi n}{L+1}.
\label{eq:kn_def}
\end{equation}
The full cavity-mediated interaction in the finite-sample eigenbasis is therefore completely specified by Eqs.~\eqref{eq:V_abgd} and \eqref{eq:Jnm_factorized}. Unlike in the translationally invariant limit, the resulting two-body vertex does not conserve pair momentum exactly; any center-of-mass momentum labeling of the Cooper channel is therefore only approximate and becomes asymptotically accurate only for states localized far from the sample boundaries or in the large-$L$ limit.

\section{Square-lattice derivation for the field of a circular loop}
\label{sec:real_loop}
Here we replace the Gaussian profile by the magnetostatic vector potential
generated by a thin circular loop of radius \(R\), centered at the origin and
lying parallel to the electronic plane at height \(z=h\).  The quantized vector
potential is written as
\begin{equation}
\hat{\bm A}(\bm r)=\bm A_{\rm loop}(\bm r;R,h)\,\hat X,
\qquad
\hat X\equiv \hat a+\hat a^\dagger ,
\label{eq:A_loop_quantized}
\end{equation}
where \(\bm r=(x,y,0)\) lies in the electron plane.  By axial symmetry,
\(\bm A_{\rm loop}\) is purely azimuthal,
\begin{equation}
\bm A_{\rm loop}(\bm r;R,h)=A_\phi(\rho;R,h)\,\hat{\bm\phi},
\quad
\rho=\sqrt{x^2+y^2},
\nonumber 
\end{equation}
with the exact filamentary-loop expression
\begin{equation}
A_\phi(\rho;R,h)
=
\frac{\mu_0 I_0}{\pi k}\sqrt{\frac{R}{\rho}}
\left[
\left(1-\frac{k^2}{2}\right)K(k)-E(k)
\right],
\nonumber 
\end{equation}
where \(K\) and \(E\) are complete elliptic integrals,
\begin{equation}
k^2=\frac{4R\rho}{(R+\rho)^2+h^2},
\nonumber 
\end{equation}
and \(I_0\) is the current scale absorbed into the mode profile.

\paragraph{Bond Peierls phases.}
Since \(h>0\), the vector potential is smooth throughout the electronic plane,
and for \(a\ll \min\{R,h\}\) the midpoint approximation remains controlled:
\begin{equation}
\Phi_{i,i+\delta}
=
\frac{e}{\hbar}\int_{\bm r_i}^{\bm r_i+\bm\delta}\bm A_{\rm loop}(\bm r;R,h)\cdot d\bm l
\simeq
\frac{ea}{\hbar}\,\bm A_{\rm loop}(\bm R;R,h)\cdot\hat{\bm\delta},
\nonumber 
\end{equation}
with bond center \(\bm R=\bm r_i+\bm\delta/2\).  Using
\(\hat{\bm\phi}=(-R_y/\rho,R_x/\rho,0)\), one finds
\begin{align}
\Phi_x(\bm R)
&\equiv \Phi_{i,i+a\hat x}
\simeq
-\frac{ea}{\hbar}\,A_\phi(\rho;R,h)\,\frac{R_y}{\rho},
\nonumber\\
\Phi_y(\bm R)
&\equiv \Phi_{i,i+a\hat y}
\simeq
+\frac{ea}{\hbar}\,A_\phi(\rho;R,h)\,\frac{R_x}{\rho}.
\label{eq:Phi_xy_loop_real}
\end{align}
Thus the bond phase remains azimuthal, but its radial dependence is now fixed by
the exact loop field rather than by a Gaussian envelope.

\paragraph{Momentum-space form factor of the loop.}
For the lattice derivation it is convenient to work in the 2D Fourier
representation over the plane \(z=0\).  The Fourier transform of the current
density of a thin loop at \(z=h\) is
\begin{align}
\bm j_{\rm loop}(\bm q)
& =
I_0 R\int_{0}^{2\pi}d\varphi\;
\hat{\bm\varphi}\,
e^{-i\bm q\cdot R\hat{\bm\rho}}\,\delta(z-h)
\nonumber 
\\
& =-2\pi i\,I_0R\,J_1(qR)\,
(\hat{\bm z}\times \hat{\bm q})\,\delta(z-h),
\nonumber 
\end{align}
where \(q=|\bm q|\).  Solving the magnetostatic equation in Coulomb gauge gives,
at the electronic plane \(z=0\), $\bm A_{\rm loop}(\bm q)$ will be 
\begin{equation}
\frac{\mu_0}{2q}\,e^{-qh}\,\bm j_{\rm loop}(\bm q)
=
-i\,\pi\mu_0 I_0R\,e^{-qh}\,
\frac{J_1(qR)}{q^2}\,
(\hat{\bm z}\times \bm q).
\nonumber 
\end{equation}
Therefore the Fourier-space Peierls phases are
\begin{align}
\Phi_x(\bm q)
&=
+i\,\Phi_R\,
e^{-qh}\,
\frac{J_1(qR)}{q^2}\,
q_y,
\nonumber\\
\Phi_y(\bm q)
&=
-i\,\Phi_R\,
e^{-qh}\,
\frac{J_1(qR)}{q^2}\,
q_x,
\label{eq:Phi_loop_q}
\end{align}
with
\begin{equation}
\Phi_R\equiv \frac{\pi\mu_0 e a I_0 R}{\hbar}.
\label{eq:PhiR_def}
\end{equation}
As before, the bond phases are odd in \(\bm q\) and purely imaginary, but the
Gaussian cutoff is replaced by the exact loop form factor
\(e^{-qh}J_1(qR)/q^2\).

\paragraph{Lattice vertex for a circular loop.}
Substituting Eq.~\eqref{eq:Phi_loop_q} into the general lattice expression for
\(g^{\rm lat}_{\bm k,\bm q}\), one obtains
\begin{align}
g^{\rm loop}_{\bm k,\bm q} 
& = \ i\,g_R\, e^{-qh}\, \frac{J_1(qR)}{q^2}\times \nonumber  \\
&  \left[ q_y\sin\!\left(k_xa+\frac{q_xa}{2}\right) - q_x\sin\!\left(k_ya+\frac{q_ya}{2}\right) \right],
\label{eq:g_loop_explicit}
\end{align}
where
\begin{equation}
g_R\equiv 2t\,\Phi_R
\label{eq:gR_def}
\end{equation}
is real.  The vertex is therefore again purely imaginary and transverse, but it
now inherits the exact nonlocal momentum structure of the loop.

In contrast to the Gaussian case, the loop profile
does not impose a Gaussian small-\(q\) cutoff; instead, the relevant scale is set
by \(R\) and the vertical separation \(h\).

\paragraph{Cavity-mediated interaction and Cooper kernel.}
Adiabatically eliminating the cavity quadrature again gives
\begin{equation}
\hat H_{\rm int}=-\frac{1}{\hbar\omega}\,\hat J^2,
\label{eq:Hint_loop}
\end{equation}
so the two-body interaction retains the same operator structure as before,
with the replacement
\(g^{\rm lat}_{\bm k,\bm q}\to g^{\rm loop}_{\bm k,\bm q}\).
Projecting onto the \(Q=0\) singlet Cooper channel,
\((\bm k\uparrow,-\bm k\downarrow)\to(\bm p\uparrow,-\bm p\downarrow)\),
with \(\bm q=\bm p-\bm k\), we obtain
\begin{equation}
V^{\rm BCS}_{\bm k,\bm p}
=
-\frac{1}{\hbar\omega}\,
g^{\rm loop}_{\bm k,\bm p-\bm k}\,
g^{\rm loop}_{-\bm k,\bm k-\bm p}.
\label{eq:VBCS_loop_def}
\end{equation}
Using Eq.~\eqref{eq:g_loop_explicit}, this becomes
\begin{align}
V^{\rm BCS}_{\bm k,\bm p}
&=
V_R\,
e^{-2|\bm p-\bm k|h}\,
\frac{J_1^2(|\bm p-\bm k|R)}{|\bm p-\bm k|^4}
\Bigg[
(p_y-k_y)\sin\!\frac{(k_x+p_x)a}{2}
\nonumber\\
&\qquad\qquad\qquad\qquad
-(p_x-k_x)\sin\!\frac{(k_y+p_y)a}{2}
\Bigg]^2,
\nonumber 
\end{align}
where
\begin{equation}
V_R\equiv \frac{g_R^2}{\hbar\omega}>0.
\nonumber 
\end{equation}
The kernel is therefore again positive semidefinite as a quadratic form, while
entering the gap equation with the usual overall minus sign.

\paragraph{Comparison to a Gaussian profile.}
Relative to the Gaussian ansatz, the circular-loop geometry replaces the
short-range factor \(e^{-q^2\ell^2/2}\) by the exact magnetostatic form
\(e^{-qh}J_1(qR)/q^2\).  The scale \(h\) is typically few nanometers of for the separation of the loop and the 2DEG so it does not suppress large momentum transfer
through the evanescent factor \(e^{-qh}\), while \(R\) imprints the oscillatory
Bessel structure associated with the loop circumference.

\section{chiral ground state from PDW component components}
\label{subsec:chiral_PDW_GL}

We consider the two symmetry-related PDW ordering vectors
\begin{equation}
\mathbf Q_x=(Q,0),
\qquad
\mathbf Q_y=(0,Q),
\label{eq:QxQy_def}
\end{equation}
together with their opposite-momentum partners.
To describe the standing-wave sector, we restrict to the reduced manifold in which the \(\pm \mathbf Q_\nu\) components within each direction are locked with equal magnitude. The gap function can then be written as
\begin{align}
\hat\Delta(\mathbf r,\mathbf k)
&=
\Big[
\eta_x\,\phi_x(\mathbf k)\cos(\mathbf Q_x\!\cdot\!\mathbf r)+
\eta_y\,\phi_y(\mathbf k)\cos(\mathbf Q_y\!\cdot\!\mathbf r)
\Big],
\label{eq:Delta_LO_xy_rewrite}
\end{align}
where \(\eta_x\) and \(\eta_y\) are complex amplitudes. For the odd-parity triplet channel selected by the transverse interaction on the square lattice, the leading basis functions are
\begin{equation}
\phi_x(\mathbf k)=\sin(k_y a),
\qquad
\phi_y(\mathbf k)=-\sin(k_x a),
\label{eq:triplet_formfactors_xy}
\end{equation}
which are exchanged by the \(C_4\) symmetry of the lattice.

The quadratic Ginzburg--Landau functional is fixed by symmetry to be
\begin{equation}
F_2
=
r\left(|\eta_x|^2+|\eta_y|^2\right),
\label{eq:F2_xy_rewrite}
\end{equation}
so that the two components become critical simultaneously at \(r=0\).
To determine the structure of the ordered state, one must therefore retain the quartic terms.

A controlled derivation follows from integrating out the fermions in the presence of the pairing field. Writing the Nambu action in the standard form, the free energy reads
\begin{equation}
F[\Delta]
=
\frac{1}{g}\int d^2r\sum_{\mathbf k} |\Delta(\mathbf r,\mathbf k)|^2
-
T\,\Tr\ln\!\left(1-G_0\Sigma\right),
\nonumber 
\end{equation}
where \(G_0\) is the normal-state Green's function and \(\Sigma\) is the anomalous self-energy generated by \(\Delta\).
Expanding the logarithm,
\begin{equation}
-\,T\,\Tr\ln(1-G_0\Sigma)
=
T\sum_{m=1}^\infty \frac{1}{m}\Tr(G_0\Sigma)^m,
\label{eq:log_expansion}
\end{equation}
and the quartic contribution is
\begin{equation}
F_4
=
\frac{T}{4}\Tr(G_0\Sigma)^4.
\label{eq:F4_trace}
\end{equation}

Projecting Eq.~\eqref{eq:F4_trace} onto the two-component manifold
\eqref{eq:Delta_LO_xy_rewrite} yields the general quartic form
\begin{equation}
F_4
=
u\left(|\eta_x|^2+|\eta_y|^2\right)^2
+
v\left(|\eta_x|^2-|\eta_y|^2\right)^2
+
\beta_2\,\bigl|\eta_x^2+\eta_y^2\bigr|^2,
\nonumber 
\end{equation}
where \(u\) and \(v\) are real coefficients and \(\beta_2\) is the phase-locking coefficient.
The last term is the only independent quartic invariant in the reduced basis that depends on the relative phase between \(\eta_x\) and \(\eta_y\).

We now derive \(\beta_2\) microscopically.
The mixed quartic term arises from the box contribution in Eq.~\eqref{eq:F4_trace} containing two insertions of the \(x\)-component and two insertions of the \(y\)-component.
Its order-parameter structure is
\begin{equation}
F_4^{xy}
=
\beta_2
\Big[
2|\eta_x|^2|\eta_y|^2
+
\eta_x^2\eta_y^{\ast 2}
+
\eta_x^{\ast 2}\eta_y^2
\Big],
\label{eq:F4_mixed_structure}
\end{equation}
which is equivalent to the invariant \(|\eta_x^2+\eta_y^2|^2\) up to terms already absorbed into \(u\).
For real basis functions \(\phi_x,\phi_y\), one finds
\begin{equation}
\beta_2
=
C\,T\sum_{\omega_n,\mathbf k}
\phi_x^2(\mathbf k)\phi_y^2(\mathbf k)\,
K_x(\mathbf k,\omega_n)\,
K_y(\mathbf k,\omega_n),
\label{eq:beta2_general}
\end{equation}
where \(C>0\) is a positive combinatorial constant and
\begin{equation}
K_\nu(\mathbf k,\omega_n)
=
G\!\left(\frac{\mathbf Q_\nu}{2}+\mathbf k,i\omega_n\right)
G\!\left(\frac{\mathbf Q_\nu}{2}-\mathbf k,-i\omega_n\right),
\qquad
\nu=x,y.
\label{eq:Knu_def}
\end{equation}

To proceed, we assume the weak-coupling symmetric-patch reduction appropriate to the LO state, namely that the dispersion in each PDW channel is inversion-symmetric about the pair center-of-mass momentum:
\begin{equation}
\xi\!\left(\frac{\mathbf Q_\nu}{2}+\mathbf k\right)
=
\xi\!\left(\frac{\mathbf Q_\nu}{2}-\mathbf k\right)
\equiv
\xi_\nu(\mathbf k).
\label{eq:symmetric_patch_condition}
\end{equation}
Then the normal-state Green's function
\begin{equation}
G(\mathbf p,i\omega_n)=\frac{1}{i\omega_n-\xi_{\mathbf p}}
\label{eq:G_def}
\end{equation}
gives
\begin{align}
K_\nu(\mathbf k,\omega_n)
&=
\frac{1}{
\bigl(i\omega_n-\xi_\nu(\mathbf k)\bigr)
\bigl(-i\omega_n-\xi_\nu(\mathbf k)\bigr)}
\nonumber\\
&=
\frac{1}{\omega_n^2+\xi_\nu^2(\mathbf k)}.
\label{eq:Knu_positive}
\end{align}
Equation~\eqref{eq:beta2_general} therefore reduces to
\begin{equation}
\beta_2
=
C\,T\sum_{\omega_n,\mathbf k}
\frac{\phi_x^2(\mathbf k)\phi_y^2(\mathbf k)}
{\left[\omega_n^2+\xi_x^2(\mathbf k)\right]
 \left[\omega_n^2+\xi_y^2(\mathbf k)\right]}.
\label{eq:beta2_manifest_positive}
\end{equation}
Since \(C>0\), \(\phi_x^2\phi_y^2\ge 0\), and each denominator is strictly positive, every term in the sum is nonnegative. Moreover, because \(\phi_x(\mathbf k)\phi_y(\mathbf k)\neq 0\) on a finite region of the Brillouin zone, the sum is strictly positive:
\begin{equation}
\beta_2>0.
\label{eq:beta2_positive_rewrite}
\end{equation}
Thus, within the weak-coupling symmetric-patch reduction, the sign of the phase-locking invariant is fixed microscopically. Equation~\eqref{eq:beta2_positive_rewrite} relies on the symmetric-patch condition
\eqref{eq:symmetric_patch_condition}. Without that reduction, the mixed quartic coefficient is still obtained from the fermionic box diagram, but its sign is not generically fixed by symmetry alone and must be computed from the full momentum-resolved loop.

It is now convenient to parametrize
\begin{equation}
\eta_x=\sqrt{\rho}\,\cos\theta\,e^{i\varphi_x},
\qquad
\eta_y=\sqrt{\rho}\,\sin\theta\,e^{i\varphi_y},
\label{eq:parametrization_rewrite}
\end{equation}
with relative phase
\begin{equation}
\delta=\varphi_y-\varphi_x.
\label{eq:delta_def_rewrite}
\end{equation}
Using the identity
\begin{equation}
\bigl|\eta_x^2+\eta_y^2\bigr|^2
=
\rho^2\Bigl[1-\sin^2(2\theta)\sin^2\delta\Bigr],
\label{eq:eta_identity_rewrite}
\end{equation}
the quartic free energy becomes
\begin{equation}
F_4
=
u\rho^2
+
v\rho^2\cos^2(2\theta)
+
\beta_2\rho^2
\Bigl[1-\sin^2(2\theta)\sin^2\delta\Bigr].
\label{eq:F4_theta_delta_rewrite}
\end{equation}

Because \(\beta_2>0\), the relative phase is minimized by maximizing \(\sin^2\delta\), hence
\begin{equation}
\delta=\pm \frac{\pi}{2}.
\label{eq:delta_minimum_rewrite}
\end{equation}
Substituting this into Eq.~\eqref{eq:F4_theta_delta_rewrite} gives
\begin{equation}
F_4
=
u\rho^2
+
\left(v+\beta_2\right)\rho^2\cos^2(2\theta).
\label{eq:F4_after_delta}
\end{equation}
Therefore, the coexistence of the two symmetry-related PDW components is favored when
\begin{equation}
v+\beta_2>0,
\label{eq:coexistence_condition}
\end{equation}
in which case the minimum occurs at equal amplitudes,
\begin{equation}
|\eta_x|=|\eta_y|,
\qquad
\theta=\frac{\pi}{4}.
\label{eq:equal_amplitude_rewrite}
\end{equation}
Combining Eqs.~\eqref{eq:delta_minimum_rewrite} and
\eqref{eq:equal_amplitude_rewrite}, the two minima are
\begin{equation}
(\eta_x,\eta_y)\propto (1,\pm i).
\label{eq:chiral_minima_rewrite}
\end{equation}

The resulting gap function is
\begin{align}
\hat\Delta_{\rm TRSB}(\mathbf r,\mathbf k)
&\propto
\bigl[i(\mathbf d\!\cdot\!\boldsymbol\sigma)\sigma^y\bigr]
\Big[
\phi_x(\mathbf k)\cos(Qx)
\nonumber\\
&
\pm i\,\phi_y(\mathbf k)\cos(Qy)
\Big].
\label{eq:Delta_chiral_final_rewrite}
\end{align}
This state is a bidirectional PDW with a relative phase \(\pm \pi/2\), and hence it spontaneously breaks time-reversal symmetry.

The chirality is characterized by the Ising order parameter
\begin{equation}
\chi
=
i\left(\eta_x\eta_y^\ast-\eta_y\eta_x^\ast\right)
=
2\,\Im\!\left(\eta_x\eta_y^\ast\right),
\label{eq:chirality_def_rewrite}
\end{equation}
which is nonzero for the two minima \eqref{eq:chiral_minima_rewrite}.
Under time reversal,
\begin{equation}
\mathcal T:\ (\eta_x,\eta_y)\mapsto(\eta_x^\ast,\eta_y^\ast),
\label{eq:TR_action_rewrite}
\end{equation}
so the two states \((1,+i)\) and \((1,-i)\) are exchanged.
We therefore conclude that, within the reduced LO manifold and for \(v+\beta_2>0\), the weak-coupling theory selects a time-reversal-breaking bidirectional PDW ground state.

\subsection{ Reduction to the chiral sector}
\label{app:four_component_PDW}

In this Appendix we formulate the quartic Ginzburg--Landau theory for the four
axial PDW components
\begin{equation}
\Delta_{x\pm}\equiv \Delta_{\pm \bm Q_x},
\qquad
\Delta_{y\pm}\equiv \Delta_{\pm \bm Q_y},
\bm Q_x=Q_0\hat x,
\bm Q_y=Q_0\hat y,
\label{eq:Qxy_app}
\nonumber
\end{equation}
and show how the reduced two-component LO theory discussed in the main text
emerges as a controlled submanifold.

We take the pairing Hamiltonian with
the odd-parity basis functions that satisfy
\begin{equation}
\phi_\mu(-\bm k)=-\phi_\mu(\bm k).
\label{eq:oddphi_app}
\end{equation}
For the leading axial triplet harmonics on the square lattice one may choose
\begin{equation}
\phi_x(\bm k)=\sin(k_y a),
\qquad
\phi_y(\bm k)=-\sin(k_x a),
\label{eq:triplet_phi_app}
\end{equation}
although the derivation below does not require this specific form until the
weak-coupling reduction is invoked.

The free-energy functional obtained after integrating out the fermions is
\begin{equation}
F[\Delta]
=
F_N
+\sum_{\alpha}\frac{|\Delta_\alpha|^2}{\lambda_\alpha}
-
T\,\Tr\ln\!\left(1-\hat G_0\hat\Sigma_\Delta\right),
\label{eq:Ffunctional_app}
\end{equation}
with \(\alpha\in\{x+,x-,y+,y-\}\) and
\begin{equation}
G_0(\bm p,i\omega_n)=\frac{1}{i\omega_n-\xi_{\bm p}}.
\label{eq:G0_app}
\end{equation}
Expanding the logarithm,
\begin{equation}
-\,T\,\Tr\ln\!\left(1-\hat G_0\hat\Sigma_\Delta\right)
=
T\sum_{m=1}^\infty \frac{1}{m}\Tr\!\left(\hat G_0\hat\Sigma_\Delta\right)^m,
\label{eq:log_expand_app}
\end{equation}
so that the quartic contribution is
\begin{equation}
F_4
=
\frac{T}{4}\Tr\!\left(\hat G_0\hat\Sigma_\Delta\right)^4.
\label{eq:F4trace_app}
\end{equation}

\subsection{Symmetry-allowed quartic invariants}

Translation symmetry requires every quartic monomial to carry zero total
crystal momentum. For the four axial components \(\Delta_{x\pm}\) and
\(\Delta_{y\pm}\), the quartic free energy therefore contains the generic
invariants
\begin{align}
F_4
=
&\;
u_1\sum_{\mu=x,y}\sum_{s=\pm} |\Delta_{\mu s}|^4
+
u_2\sum_{\mu=x,y} |\Delta_{\mu +}|^2 |\Delta_{\mu -}|^2
\nonumber\\
&+
u_3\left(|\Delta_{x+}|^2+|\Delta_{x-}|^2\right)
   \left(|\Delta_{y+}|^2+|\Delta_{y-}|^2\right)
\nonumber\\
&+
u_4\Bigl(
\Delta_{x+}\Delta_{x-}\Delta_{y+}^\ast\Delta_{y-}^\ast
+\mathrm{c.c.}
\Bigr)
+\cdots ,
\label{eq:F4_general_fourcomp}
\end{align}
where the last term denotes additional symmetry-allowed quartic terms that are
not phase sensitive in the reduced LO sector. The key phase-locking invariant is
the last term in Eq.~\eqref{eq:F4_general_fourcomp}. It is allowed because
\begin{equation}
\bm Q_x+(-\bm Q_x)-\bm Q_y-(-\bm Q_y)=0.
\label{eq:momentum_conservation_u4}
\end{equation}

\subsection{Exact microscopic definition of \(u_4\)}

The coefficient \(u_4\) is defined microscopically by projecting the quartic
trace \eqref{eq:F4trace_app} onto the monomial
\(\Delta_{x+}\Delta_{x-}\Delta_{y+}^\ast\Delta_{y-}^\ast\):
\begin{equation}
u_4
=
\frac{T}{4}
\left.
\frac{\partial^4}{\partial \Delta_{x+}\,\partial \Delta_{x-}\,
\partial \Delta_{y+}^\ast\,\partial \Delta_{y-}^\ast}
\Tr\!\left(\hat G_0\hat\Sigma_\Delta\right)^4
\right|_{\Delta=0}.
\nonumber 
\end{equation}
Equivalently,
\begin{equation}
u_4
=
T\sum_{\omega_n}\frac{1}{N}\sum_{\bm k}\,
\mathcal B_{xy}(\bm k,i\omega_n),
\label{eq:u4_boxkernel_app}
\end{equation}
where \(\mathcal B_{xy}\) is the properly momentum-routed fermionic box kernel,
including the sum over inequivalent cyclic orderings of the four pairing
vertices that contribute to the monomial
\(\Delta_{x+}\Delta_{x-}\Delta_{y+}^\ast\Delta_{y-}^\ast\).

At the full lattice level, \(\mathcal B_{xy}\) depends on shifted internal
momenta and on form factors evaluated at different arguments. In particular,
the exact box does \emph{not} generically collapse to a single common-momentum
product of the form
\begin{equation}
\phi_x^2(\bm k)\phi_y^2(\bm k)\,
G_{x+}\bar G_{x-}G_{y+}\bar G_{y-},
\label{eq:not_exact_box_app}
\end{equation}
and the sign of \(u_4\) is therefore not fixed by symmetry alone in the full
lattice theory. Away from the controlled weak-coupling reduction discussed
below, \(u_4\) must be obtained by evaluating the full box kernel
\(\mathcal B_{xy}\).

\subsection{Weak-coupling symmetric-patch reduction}

A transparent analytic result follows in the weak-coupling regime, where each
PDW component is concentrated near a symmetry-related patch and the relevant
pairing momenta satisfy
\begin{equation}
\bm Q_x=2\bm K_x,
\qquad
\bm Q_y=2\bm K_y.
\label{eq:Qequal2K_app}
\end{equation}
Writing the internal momentum relative to the corresponding patch center as
\(\bm q\), we assume that the dispersion is inversion symmetric within each
paired patch:
\begin{equation}
\xi\!\left(\frac{\bm Q_\nu}{2}+\bm q\right)
=
\xi\!\left(\frac{\bm Q_\nu}{2}-\bm q\right)
\equiv
\varepsilon_\nu(\bm q),
\qquad
\nu=x,y.
\label{eq:symmetric_patch_app}
\end{equation}
In the same approximation the form factors may be treated as smooth patch
functions \(\phi_\nu(\bm q)\), and the inequivalent box routings reduce to the
same positive kernel up to an overall combinatorial constant \(C>0\). One then
obtains
\begin{equation}
u_4
\approx
C\,T\sum_{\omega_n}\frac{1}{N}\sum_{\bm q}
\frac{\phi_x^2(\bm q)\phi_y^2(\bm q)}
{\bigl(\omega_n^2+\varepsilon_x^2(\bm q)\bigr)
 \bigl(\omega_n^2+\varepsilon_y^2(\bm q)\bigr)}.
\label{eq:u4_patch_app}
\end{equation}
For the axial triplet lattice harmonics,
\begin{equation}
\phi_x(\bm q)\sim \sin(k_y a),
\qquad
\phi_y(\bm q)\sim \sin(k_x a),
\label{eq:phi_patch_app}
\end{equation}
but the sign argument below requires only that \(\phi_x\) and \(\phi_y\) be
real, so that \(\phi_x^2\phi_y^2\ge 0\).

The Matsubara sum is elementary:
\begin{align}
T\sum_{\omega_n}
\frac{1}{(\omega_n^2+\varepsilon_x^2)(\omega_n^2+\varepsilon_y^2)}
\nonumber
&=
\frac{1}{\varepsilon_y^2-\varepsilon_x^2}
\left[
\frac{\tanh(\varepsilon_x/2T)}{2\varepsilon_x} \right.
\\ \left. -
\frac{\tanh(\varepsilon_y/2T)}{2\varepsilon_y}
\right]
&=
\frac{f(\varepsilon_x)-f(\varepsilon_y)}
{2(\varepsilon_y^2-\varepsilon_x^2)},
\label{eq:freqsum_correct_app}
\end{align}
where
\begin{equation}
f(\varepsilon)\equiv \frac{\tanh(\varepsilon/2T)}{\varepsilon}.
\label{eq:fdef_app}
\end{equation}
Since \(f(\varepsilon)\) is strictly decreasing for \(\varepsilon>0\), the
right-hand side of Eq.~\eqref{eq:freqsum_correct_app} is nonnegative.
Therefore every term in Eq.~\eqref{eq:u4_patch_app} is nonnegative, and one
finds
\begin{equation}
u_4>0
\label{eq:u4positive_app}
\end{equation}
\text{within the weak-coupling symmetric-patch approximation.}
\subsection{Phase locking and reduction to the LO sector}

Writing
\begin{equation}
\Delta_{\mu\pm}=|\Delta_{\mu\pm}|e^{i\theta_{\mu\pm}},
\label{eq:Deltaphase_app}
\end{equation}
the phase-sensitive quartic contribution is
\begin{equation}
F_{u_4}
=
2u_4\,
|\Delta_{x+}\Delta_{x-}\Delta_{y+}\Delta_{y-}|
\cos\Phi,
\Phi\equiv
\theta_{x+}+\theta_{x-}-\theta_{y+}-\theta_{y-}.
\label{eq:Fu4_phase_app}
\end{equation}
Hence, if \(u_4>0\), the minimum occurs at
\begin{equation}
\Phi=\pi
\qquad (\mathrm{mod}\;2\pi).
\label{eq:phase_locking_app}
\end{equation}

We now restrict to the LO submanifold, in which the \(\pm \bm Q_\nu\) components
within each direction are locked with equal magnitude and equal phase:
\begin{equation}
\Delta_{x+}=\Delta_{x-}\equiv \frac{\eta_x}{2},
\qquad
\Delta_{y+}=\Delta_{y-}\equiv \frac{\eta_y}{2}.
\label{eq:LO_restriction_app}
\end{equation}
The real-space gap function then becomes
\begin{equation}
\Delta(\bm r,\bm k)
=
\eta_x\,\phi_x(\bm k)\cos(\bm Q_x\!\cdot\!\bm r)
+
\eta_y\,\phi_y(\bm k)\cos(\bm Q_y\!\cdot\!\bm r).
\label{eq:LO_gap_app}
\end{equation}
Substituting Eq.~\eqref{eq:LO_restriction_app} into
Eq.~\eqref{eq:Fu4_phase_app} gives
\begin{equation}
F_{u_4}^{\rm LO}
=
\frac{u_4}{8}\,
|\eta_x|^2|\eta_y|^2
\cos\!\bigl[2(\varphi_x-\varphi_y)\bigr],
\label{eq:Fu4_LO_app}
\end{equation}
where \(\eta_\nu=|\eta_\nu|e^{i\varphi_\nu}\).
For \(u_4>0\), Eq.~\eqref{eq:Fu4_LO_app} is minimized by
\begin{equation}
\varphi_y-\varphi_x=\pm \frac{\pi}{2}.
\label{eq:relative_phase_LO_app}
\end{equation}
Thus the four-component quartic invariant reduces in the LO sector to the
time-reversal-breaking chiral combination
\begin{equation}
(\eta_x,\eta_y)\propto (1,\pm i).
\label{eq:chiral_LO_app}
\end{equation}

Up to a trivial normalization factor, the phase-locking coefficient in the
reduced two-component LO theory is therefore
\begin{equation}
\beta_2=\frac{u_4}{8},
\label{eq:beta2_u4_relation_app}
\end{equation}
so that the positivity of \(u_4\) in the weak-coupling symmetric-patch
approximation is equivalent to the positivity of \(\beta_2\) in the main text.

\bibliography{References}

\end{document}